\documentclass{aa} 
\usepackage{graphicx}
\usepackage{txfonts}
\usepackage{xcolor}
\usepackage[flushleft]{threeparttable}
\begin{document} 
   \title{Universal bolometric corrections for AGN over 7 luminosity decades}

\author{F. Duras
          \inst{1,2,3}
          \and
          A. Bongiorno \inst{2}
          \and
          F. Ricci \inst{4}
          \and
           E. Piconcelli \inst{2}
          \and
          F. Shankar \inst{5}
          \and
          E. Lusso \inst{6,7}
          \and 
          S. Bianchi \inst {1}
          \and
          F. Fiore \inst {8}
          \and
          R. Maiolino \inst {9}
          \and 
          A. Marconi \inst{6}
          \and
          F. Onori \inst {10}
          \and 
          E. Sani \inst {11}
          \and 
          R. Schneider \inst {12,2}
          \and 
          C. Vignali \inst{13,14}
          \and 
          F. La Franca \inst {1}
            }

   \institute{Dipartimento di Matematica e Fisica, Universit\`a Roma Tre, via della Vasca Navale 84, I-00146, Roma, Italy
            \and
             Osservatorio Astronomico di Roma -INAF, via Frascati 33, 00040 Monteporzio Catone, Italy
           \and
           Aix Marseille Univ, CNRS, CNES, LAM, Marseille, France
           \and
           Instituto de Astrof\'isica and Centro de Astroingenier\'ia,
            Facultad de F\'isica, Pontificia Universidad Cat\'olica de Chile,
            Casilla 306, Santiago 22, Chile
            \and
            Department of Physics and Astronomy, University of Southampton, Highfield, SO17 1BJ, UK
            \and
           Dipartimento di Fisica e Astronomia, Universit\`a di Firenze, via G. Sansone 1, 50019 Sesto Fiorentino, Firenze, Italy
           \and
            Osservatorio Astrofisico di Arcetri - INAF, 50125 Firenze, Italy
            \and
            Osservatorio Astronomico di Trieste - INAF, via Tiepolo 11, I-34143 Trieste, Italy
            INAF - Osservatorio Astronomico di Trieste, via G. Tiepolo 11, I-34124 Trieste, Italy
            \and
            Cavendish Laboratory, University of Cambridge, 19 J. J. Thomson Ave., Cambridge CB3 0HE, UK
            \and
            Istituto di Astrofisica e Planetologia Spaziali (INAF), via del Fosso del Cavaliere 100, Roma, I-00133, Italy
            \and
            European Southern Observatory, Santiago, Chile
            \and
            Dipartimento di Fisica, Universit\`a di Roma La Sapienza, I-00185 Roma, Italy
            \and
            Dipartimento di Fisica e Astronomia, Alma Mater Studiorum, Universit\`a degli Studi di Bologna, Via Gobetti 93/2, I-40129 Bologna, Italy
            \and
            Osservatorio Astronomico di Bologna -INAF, via Ranzani 1, I-40127 Bologna, Italy
           }
   \date{Received October 1,2019; accepted January 24,2020}
% \abstract{}{}{}{}{} 
% 5 {} token are mandatory
 
  \abstract
   {The AGN bolometric correction is a key element to understand BH demographics and compute accurate BH accretion histories from AGN luminosities. However, current estimates still differ from each other by up to a factor of two to three, and rely on extrapolations at the lowest and highest luminosities. }
   {Here we revisit this fundamental issue presenting general hard X-ray ($K_{X}$) and optical ($K_{O}$) bolometric corrections, computed combining several AGN samples spanning the widest (about 7 dex) luminosity range ever used for this kind of studies.}
   {We analysed a total of $\sim 1000$ type 1 and type 2 AGN for which a dedicated SED-fitting has been carried out.}
   {We provide a bolometric correction separately for type 1 and type 2 AGN; the two bolometric corrections results to be in agreement in the overlapping luminosity range and therefore, for the first time, a universal bolometric correction for the whole AGN sample (both type 1 and type 2) has been computed. We found that $K_{X}$ is fairly constant at $log(L_{BOL}/L_{\odot}) < 11$, while it increases up to about one order of magnitude at $log(L_{BOL}/L_{\odot}) \sim 14.5$. A similar increasing trend has been observed when its dependence on either the Eddington ratio or the BH mass is considered, while no dependence on redshift up to $z\sim3.5$ has been found. On the contrary, the optical bolometric correction appears to be fairly constant (i.e. $K_{O} \sim 5$) whatever is the independent variable. We also verified that our bolometric corrections correctly predict the AGN bolometric luminosity functions. According to this analysis, our bolometric corrections can be applied to the whole AGN population in a wide range of luminosity and redshift.}
   {}

   \keywords{galaxies: active -- X-rays: galaxies -- galaxies: fundamental parameters -- quasars: general
               }

   \maketitle
%
%-------------------------------------------------------------------

\section{Introduction}

   To properly study the structure and evolution of the Active Galactic Nuclei (AGN) and understand how the observed local correlations between the Super Massive Black Hole (SMBH) mass and the host galaxy (\citealt{Magorrian1998, Tremaine2002, Marconi2003}; but see also \citealt{Shankar2016} and \citealt{Shankar2019}) originate (in the AGN-galaxy co-evolution scenarios), an accurate knowledge of the AGN internal mechanisms, physical properties and then energetic budget is necessary. 

As a result, the bolometric luminosity, which is the total luminosity emitted at any wavelength by the AGN, becomes one of  the  key  parameters  to  know  with  as  much  high  accuracy  as possible.  

However, unless a wealth of multi-photometric data are available, to derive the AGN bolometric luminosity, it is necessary to use the so-called bolometric correction, which  is  defined  as  the  ratio between the bolometric luminosity and the luminosity in a given spectral  band, i.e. $K_{band}=L_{BOL}/L_{band}$.

Empirical  bolometric  corrections  have  a  strong  impact  on  our understanding of BH demographics and AGN output, being widely adopted by the scientific community in both observational (e.g. to measure  the  fraction  of  total  bolometric  emission  in  AGN  for  a given  band  luminosity)  and  theoretical studies  \citep[e.g.   in   cosmological   hydro-dynamical   simulations   to produce realistic synthetic catalogues of AGN,][]{Koulouridis2018}. Moreover, a correct knowledge of the bolometric correction is fundamental to convert the AGN luminosity function into the accretion rate history \citep{Salucci1999, Ueda2003, Volonteri2003, Shankar2004, Marconi2004, Vittorini2005, Lafranca2005, Aversa2015} and in turn derive the local SMBHs \citep[][]{Soltan1982} or, more accurately, map the coeval growth
of galaxies and their central SMBHs \citep{Bongiorno2016, Aird2018}.

The interest in this type of studies has been steadily increasing over time due to the availability of larger AGN samples. However, the characterization of the AGN bolometric luminosity is a challenging issue for several reasons. First of all, we need to take into account the contamination from the host galaxy to the nuclear emission, which results to be a source of uncertainty particularly relevant for the lower luminosity AGN. In addition, considering that the emission in different bands originates in diverse physical structures (a sub-parsec accretion disk in the UV-optical, a dusty parsec-scale torus in the infrared and kiloparsec-scale relativistic jets in the radio), each spectral region varies with distinctive timescales. As a further problem, one should also consider the anisotropy of the AGN emission, which may depend on wavelength and which may be peculiar to each specific source. Finally, given the multi-wavelength feature of the AGN emission, observations from a variety of telescopes are necessary in order to build up complete and detailed Spectral Energy Distributions (SEDs). \\

For these reasons, in the last decades several attempts have been made to accurately study the shape of the AGN SED \citep{Sanders1989, Elvis1994, Richards2006}. A key observational parameter is the optical-to-X-ray spectral index $\alpha_{ox}$, defined as the slope of a power law connecting two regions of the SED, namely the UV (i.e. the accretion disc) and the X-rays (i.e. the so-called X-ray corona), and parametrized as $\alpha_{ox} = -\frac{log(L_{2 keV}/L_{2500\AA})}{2.605}$ \citep[][]{Tananbaum1979}. Some authors have investigated the evolution of the $\alpha_{OX}$ spectral index \citep[see e.g.,][]{Kelly2008, Vasudevan2009, Lusso2010, Marchese2012, Lusso2017}, which should provide a hint about the nature of the energy generation mechanism in AGN. Many of them report that the UV-to-X-ray SED shows no significant dependence on redshift, while the primary dependence is on the UV luminosity \citep[][]{Vignali2003, Steffen2006, Lusso2016}. This supports a scenario in which the mechanism responsible for the nuclear emission of the AGN remains identical at any redshift, while it correlates with the AGN luminosity: in particular, the ratio between the X-ray luminosity and the UV/optical one decreases with increasing luminosity \citep[see e.g.][]{Martocchia2017}.\\
One of the first attempts to derive the bolometric correction has been carried out by \citet{Elvis1994}, who have estimated the bolometric corrections in different bands, using a mean energy distribution from a sample of 47 AGN, including the IR emission in the bolometric luminosity and therefore counting part of the emission twice. \citet{Shankar2004} gave a preliminary estimate of the hard X-ray bolometric correction, while \citet{Marconi2004} derived an estimate of the bolometric corrections in the optical B, soft and hard X-ray and IR bands, similarly to \citet{Hopkins2007} and \citet[][]{Shen2020}, who however included the IR emission in the computation of the AGN total luminosity.  \citet{Vasudevan2007} studied a sample of 54 X-ray bright ($L_{X} > 10^{43}$ erg/s) AGN, pointing out that particular classes of sources (as Radio-loud, X-ray weak or Narrow Line Seyferts 1) might have different bolometric correction relations if compared with the rest of the AGN population. \citet{Lusso2012} analyzed a sample of about 900 X-ray selected AGN deriving the bolometric corrections in the same bands as Marconi, separately for type 1 and type 2 sources, and valid for approximately three orders of luminosity. \citet{Runnoe2012} sampled the optical band giving the bolometric corrections at three different wavelengths (1450 $\AA$, 3500 $\AA$ and 4000 $\AA$), while IR bolometric corrections are provided in \citet{Runnoe2012b}. \citet{Krawczyk2013} focused on the characterization of the bolometric corrections of luminous broad-line AGN in a wide range of redshift ($0<z<6$). Very recently, \citet{Netzer2019} provided simple power-law approximations of the bolometric corrections in the optical and hard X-ray bands, obtained by combining  theoretical calculations of optically thick, geometrically thin accretion disks,
and observational X-ray properties of AGN. All these works, although giving good estimate of the AGN bolometric corrections in different bands, are limited by the narrow range of luminosity sampled. This implies that for both high and low luminosity objects most of the studies had to rely on extrapolations of the bolometric correction. Moreover, all of them, but \citet{Lusso2012}, focused on luminous, unobscured (type 1), sources, where the bolometric luminosity is much more easily measured, being the SED less affected by obscuration and galaxy contamination than in type 2 sources.\\
In this paper we focus on the characterization of the hard X-ray ($K_{X} = L_{BOL}/L_{X}$) and the optical ($K_{O} = L_{BOL}/L_{O}$) bolometric corrections, considering a sample spanning the widest range of luminosity ($\sim$ 7 decades) so far, and which comprises both type 1 and type 2 sources, up to $z \sim 4$. Throughout the paper $L_{X}$ is the 2-10 keV intrinsic X-ray luminosity and $L_{O}$ is the B-band 4400 $\AA$ luminosity.  \\
 \\  
The paper is organized as follows: in Section \ref{sec:sample} we present the data sample, while Section \ref{sec:seds} provides a detailed explanation of the SED-fitting method we used to disentangle the AGN and host galaxy emission components. In Section \ref{sec:kbols} we analyze the bolometric correction as a function of luminosity, i.e. Section \ref{sec:kbolx} presents the results of the new hard X-ray bolometric correction for the type 1 and the type 2 sample separately, and then a general relation for the whole population; while Section \ref{sec:kbolo} is dedicated to the optical bolometric correction for type 1 AGN. In Section \ref{sec:kbolz} we study the bolometric correction as a function of redshift. Section \ref{sec:ledd} is dedicated to the study of the dependence of the bolometric corrections on the Black Hole mass and Eddington ratio, while in Section \ref{sec:fdilumin} we derive the AGN unobscured bolometric luminosity function using the newly computed bolometric corrections. Finally, Section \ref{sec:disc} and \ref{sec:concl} present the discussion and the summary of the work.\\
In what follows, we adopt a $ \Lambda$CDM cosmology with $ H_0 = 70$, $ \Omega_m = 0.30$, $\Omega_{\Lambda} = 0.70$. All the uncertainties and spreads, unless otherwise stated, are quoted at 68$\%$ (1$\sigma$) confidence level.

\section{Data sample}

\begin{figure}
\includegraphics[scale=0.45]{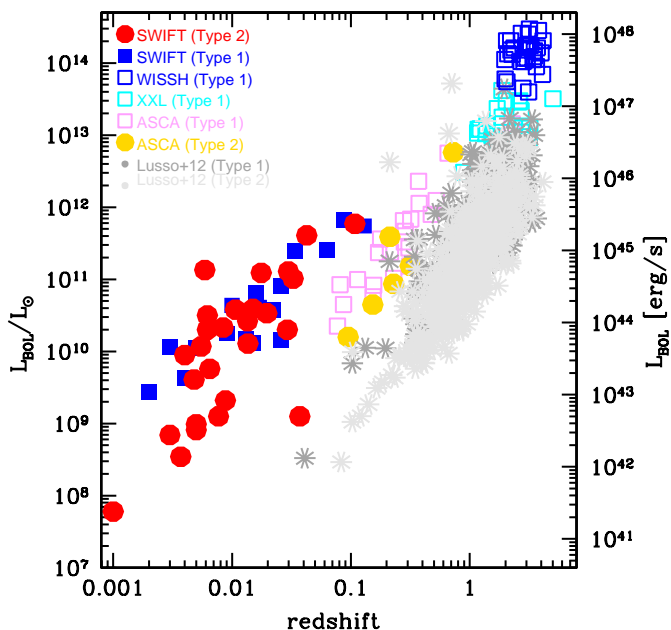}
\caption{Intrinsic X-ray luminosity in the 2-10 keV band as a function of the redshift of the SWIFT type 1 (blue filled squares) and type 2 (red filled circles) samples, the X-WISSH sample (blue open squares), the COSMOS type 1 (grey asterisks) and type 2 (lightgrey asterisks) sources, the ASCA type 1 (pink open squares) and type 2 (gold filled circles) objects and the XXL type 1 (cyan open squares) AGN.}
\label{fig:distrib}
\end{figure}

\label{sec:sample}
We used a collection of $\sim$1000 sources belonging to five different AGN samples in order to cover the widest luminosity range ($\sim$ 7 dex) ever sampled for this kind of studies (Figure \ref{fig:distrib}).\\ 

For this work we considered only radio-quiet AGN. We have therefore excluded from the analysis all sources with logL(1.4 GHz [W/Hz]) > 26  \citep[see e.g.][]{Miller1990,Lafranca2010}. 

\subsection*{The SWIFT sample}

The low-luminosity end is populated by the SWIFT sample, which consists of a collection of 21 type 1 ($2 \times 10^{41}$ erg/s < $L_{X}$ < $8 \times 10^{45}$ erg/s)  and 27 type  2  ($3 \times 10^{40}$  erg/s < $L_{X}$ < $3 \times 10^{44}$ erg/s) AGN\footnote{With the term type 1 AGN we define those targets showing broad emission lines in the optical spectra, while we refer to type 2 AGN as those AGN where there is no (Seyfert 2) or weak (intermediate $1.8 - 1.9$) evidence of the broad line region, or even no lines at all, in their optical spectra.} at $z<0.16$, from the Swift/BAT 70-month catalogue, which is considered one of the most complete hard X-ray surveys, containing about 1200 hard X-ray objects in the 14-195 keV band, of which about 700 are classified as AGN sources \citep{Baumgartner2013}. For these AGN, the intrinsic hard X-ray luminosity is provided by \citet{Riccic2017}. \\
\noindent
The SWIFT type 2 AGN belong to an original sample of 41 AGN, for which our group obtained near-infrared spectroscopic observations using ISAAC and X-Shooter at VLT, and LUCI at LBT \citep{Onori1}. Six additional sources with data from literature have been added in a second time. Among this total sample of 47 AGN, a fraction of $\sim$30\%, after deep NIR spectroscopy, showed faint broad emission lines that allowed to estimate their BH mass, $M_{BH}$, \citep[see][]{Onori2} using the BH mass virial estimator calibrated by \citet{Ricci1}.\\
\noindent The SWIFT type 1 AGN belong to a original sample of 33 sources presented in \citet{Ricci2} and included in the Swift/BAT 70-month catalogue, whose $M_{BH}$ have been measured via reverberation mapping techniques, and for which reliable bulge classification is available \citep{Ho2014}\footnote{This sample has been adopted by \citet{Ricci1} to calibrate a virial BH mass estimator that can be used to measure $M_{BH}$ of low-luminosity obscured AGN.}. 
According to \citet{Onori2} and \citet{Ricci2}, both the local type 1 and the type 2 sources can be considered representative of the AGN populations included in the Swift/BAT catalogue. As shown in Figure \ref{fig:distrib}, this sample allows to probe the lowest luminosity range of our entire data-set.\\
In order to estimate the physical properties of the SWIFT AGN sample, a multi-wavelength photometric data-set has been collected. We removed two RL sources from the original Swift/BAT 70-month catalogue and we considered only those data whose photometric aperture was representative of the flux of the entire galaxy. We then collected photometric data for a total of 36 bands, using the NED database\footnote{https://ned.ipac.caltech.edu/} and other public catalogues: far and near UV data are from GALEX observations \citep{Gil2007}; U, B, V, R  and I bands are taken from the Atlas of galaxies by \citet{Devauc1991} or from the SDSS DR12\footnote{https://www.sdss.org/dr12/}; the near-IR J,H and K bands from 2MASS\footnote{https://irsa.ipac.caltech.edu/Missions/2mass.html}; data at 3 $\mu m$, 4.5 $\mu m$, 12 $\mu m$ and 22 $\mu m$ are from WISE\footnote{https://irsa.ipac.caltech.edu/Missions/wise.html}; data at 12 $\mu m$, 25 $\mu m$, 60 $\mu m$ and 100 $\mu m$ are from IRAS and from the IRAS Revised Bright Galaxy Sample by \citet{Sanders2003}; 
120 $\mu m$, 150 $\mu m$, 170 $\mu m$, 180 $\mu m$ and 200 $\mu m$- band are from ISOPHOT \citep{Spinoglio2002}; finally, FIR data at 250 $\mu m$, 350 $\mu m$ and 500 $\mu m$  are from Herschel \citep{Shimizu2016} or BLAST \citep{Wiebe2009}. We discarded those sources for which the data were not sufficient to guarantee a good photometric coverage. In particular, we included only the objects for which: (1) we have more than three bands for each region of the spectrum, i.e. UV/optical, near-IR and mid-IR; and (2) each photometric value has a signal-to-noise ratio $S/N$ > 5. For this reason, 20 type 2 sources out of 46, and 12 type 1 sources out of 32, were excluded from our analysis. Therefore the local sample of SWIFT AGN includes 26 type 2 and 20 type 1 AGN. This selection could induce a bias in the SEDs of the sources analysed. We 
investigated this possibility by studying the X-ray/MIR ratio distribution (using the WISE W1 band, whose value is available for all the objects). It results that the sample of the excluded sources (OUT sample) shows an average X-ray/MIR ratio 0.2 dex larger than the included ones (IN sample) at 2.5 $\sigma$, while in the total sample the X-ray/MIR ratio is 0.09 dex larger than the IN sample. However, as discussed later, this bias has a negligible effects on the final derived bolometric correction relations and on the main conclusions of this paper (see Section \ref{sec:kbolx}). 

\subsection*{The X-WISSH sample}
At the brightest end, we started from the WISSH (WIse Selected Hyperluminous) sample \citep[see ][]{Bischetti2017}, composed of 86 hyper-luminous ($L_{BOL}$ >  2 $\times 10^{47}$ erg/s) type 1 AGN in the redshift range $2<z<4$, detected in the MIR by WISE. 41 out of the 86 sources have been observed in the X-rays by XMM or Chandra \citep[see][]{Martocchia2017}. Six X-ray undetected and two RL sources were excluded from our analysis, resulting in a final X-WISSH sample of 33 type 1 AGN. For all of them, a wide photometric coverage (from the X-ray to the MIR) is available.

\noindent
\subsection*{The COSMOS sample}
This sample includes 369 type 1 and 484 type 2 sources from the XMM-COSMOS catalogue studied by \citet[][{hereafter L12}]{Lusso2012}, in the redshift range $0.1 < z < 4$ and with average hard X-ray luminosity of $L_{X} \sim 10^{44}$ erg/s. From the sample presented in L12, we rejected 4 RL type 1 and 3 RL type 2 sources \footnote{The original type 2 sample in L12 was composed of 488 AGN, but one source had a wrong spectroscopic classification and was eventually classified as an inactive galaxy.}. The intrinsic hard X-ray luminosity of the sources are published in L12, and we refer to that work for the derivation of their physical properties. \\

\noindent
\subsection*{The ASCA sample}
This sample belongs to the ASCA Medium Sensitivity Survey in the northern sky (AMSSn) sample analyzed by \citet[][]{Akiyama2003}. We used all the AGN sources for which good photometric data were available (see the criteria adopted for the SWIFT sample). Their photometry has been obtained by cross-correlating data from the SDSS, the 2MASS survey and WISE. The final ASCA sample, after removing  one type 1 RL AGN, results in 22 type 1 and 6 type 2 sources, up to $z \sim 2.3$. The original X-ray luminosity published in Table 3 of \citet[][]{Akiyama2003}, have been re-scaled according to our adopted values of the cosmological parameters. \\

\subsection*{The XXL sample}
Finally, we used a sub-sample of AGN from the XXL-N survey \citep{Pierre2016}. From the 8445 point-like X-ray sources presented in the work by \citet{Liu2016}, we selected those AGN with 2-10 keV photon counts greater than 50. Among them, we chose the ones with high bolometric luminosities ($L_{BOL} > 10^{46.5}$ erg/s, as listed in their Table 2) in order to achieve better statistical significance in the highest luminosity regime just below the one probed by the X-WISSH sample. We removed 6 RL sources and therefore the final XXL sample  consists of 31 type 1 AGN in the redshift range $0.9 < z < 5$. \\

In all the above samples, when available, the 2-10 keV intrinsic X-ray luminosities derived from X-ray spectral fitting were used. Otherwise, they have been re-scaled by assuming a typical common $\Gamma$=1.8 photon index value.\\
We stress that all the described samples but the X-WISSH one are X-ray selected, which could in principle introduce a bias in the estimate of the bolometric correction. However, we will show in Section \ref{sec:kbolx} that such effect can be considered negligible.

\section{Spectral Energy Distribution}
\label{sec:seds}

In this Section we focus on the SED characterization and on the consequent derivation of the main physical properties (in particular the bolometric luminosity) of the sources belonging to the SWIFT, X-WISSH, ASCA and XXL samples. 
The properties of the COSMOS sample have been taken from L12, who used an SED-fitting method similar to the one adopted in this paper (see Section \ref{sec:resul}).

\subsection{SED-fitting}
The SED fitting procedure is based on a modified version of the code described in \citet{Duras2017} (hereafter D17). The observed SED of each source is the result of different components produced by both the nuclear engine and the stellar light \citep[e.g.,][]{Bongiorno2012,Berta2016,Rivera2016}. The nuclear emission produces two bumps in the UV and NIR regimes which create a dip at around 1$\mu$m \citep{Sanders1989, Elvis1994, Richards2006}. The UV bump is due to thermal emission from the accretion disc \citep{Czerny1987}, while the NIR bump to the re-emission at longer wavelengths of the intrinsic primary radiation absorbed by hot dusty clouds in the torus. In the same way, the stellar light produces  both a direct UV/optical emission (mainly due to hot stars) and a MIR-FIR component (ascribed to its reprocessing by the cold dust located on galactic scales). How much these two components contribute to the global SED depends on a number of factors, the most important ones being their relative luminosity and the level of obscuration affecting each of them.  \\

\begin{figure}
    \centering
    \includegraphics[scale=0.45]{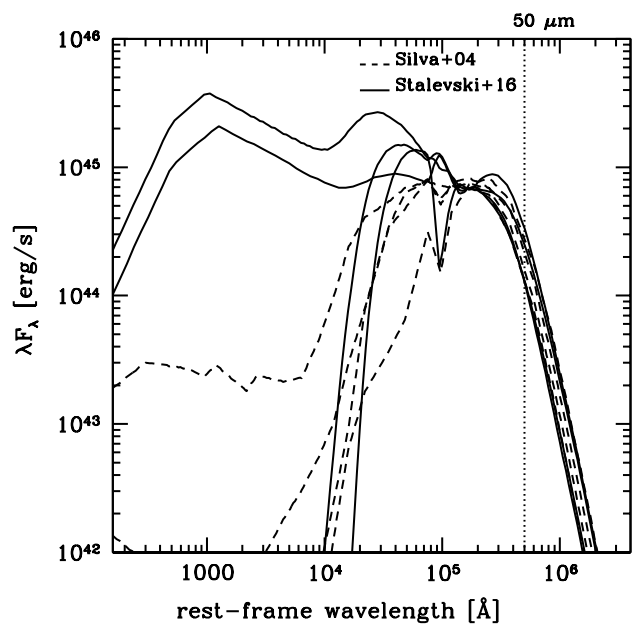}
    \caption{Examples of UV-to-IR AGN templates with different level of absorption, from \citet{Silva2004} and \citet{Stalevski2016}. The vertical dotted line at $\sim 50 \mu m$ shows the position of the pivotal photometric point derived from the X-ray luminosity, where the AGN templates are very similar.}
    \label{fig:agn}
\end{figure}

\begin{figure*}
\centering
\includegraphics*[scale=0.34]{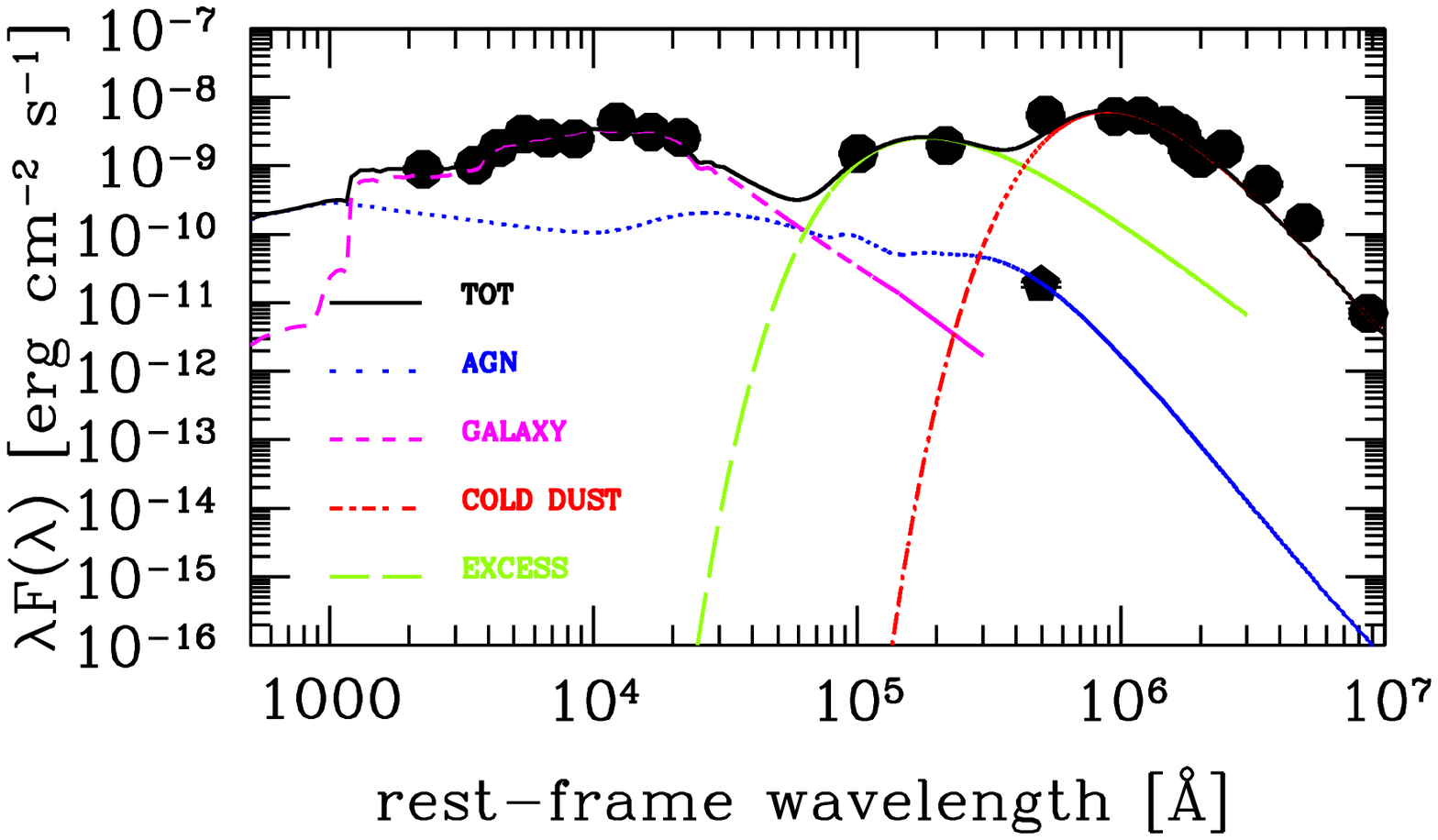} \,
\includegraphics*[scale=0.34]{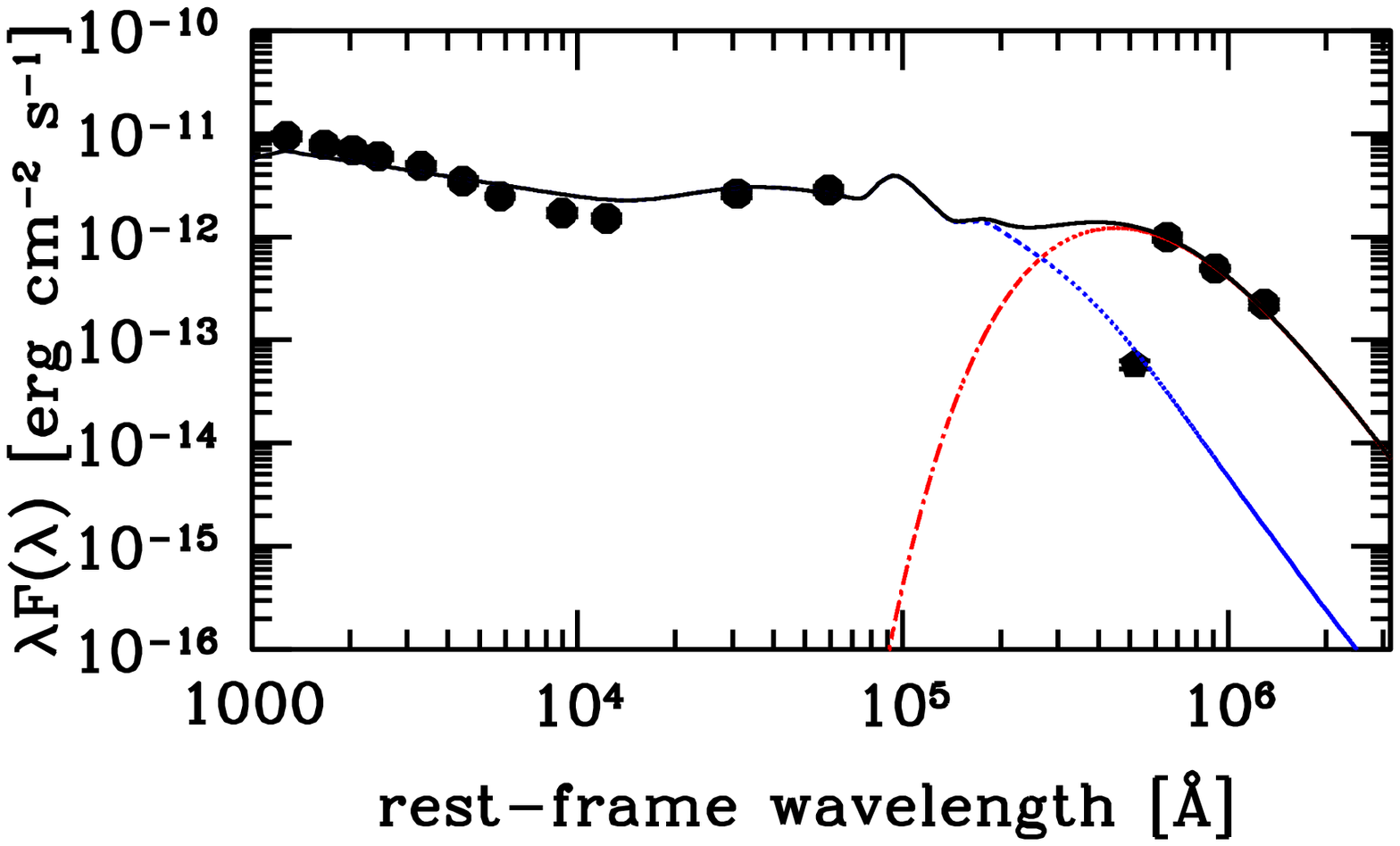}\\
\includegraphics*[scale=0.34]{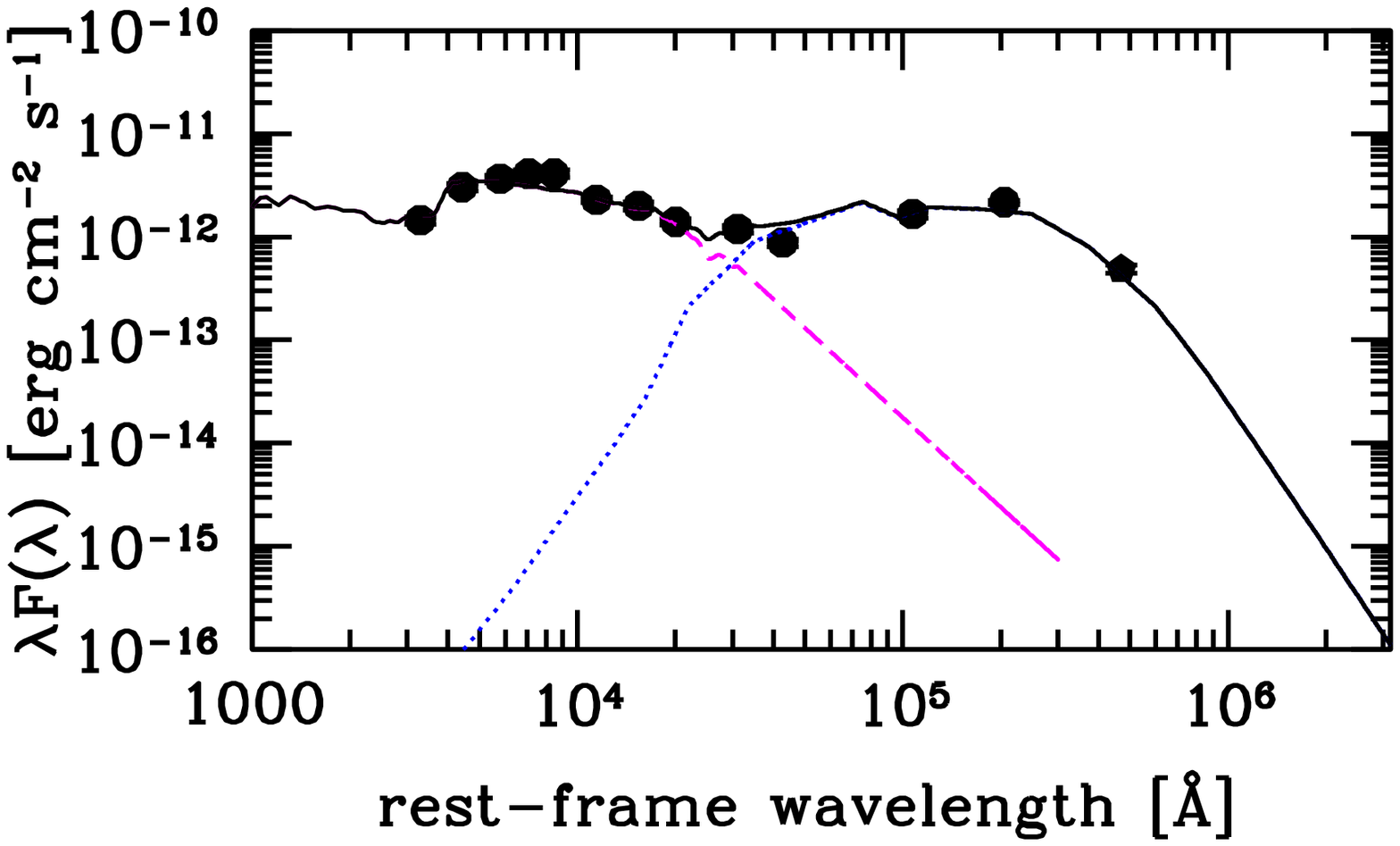}\,
\includegraphics*[scale=0.34]{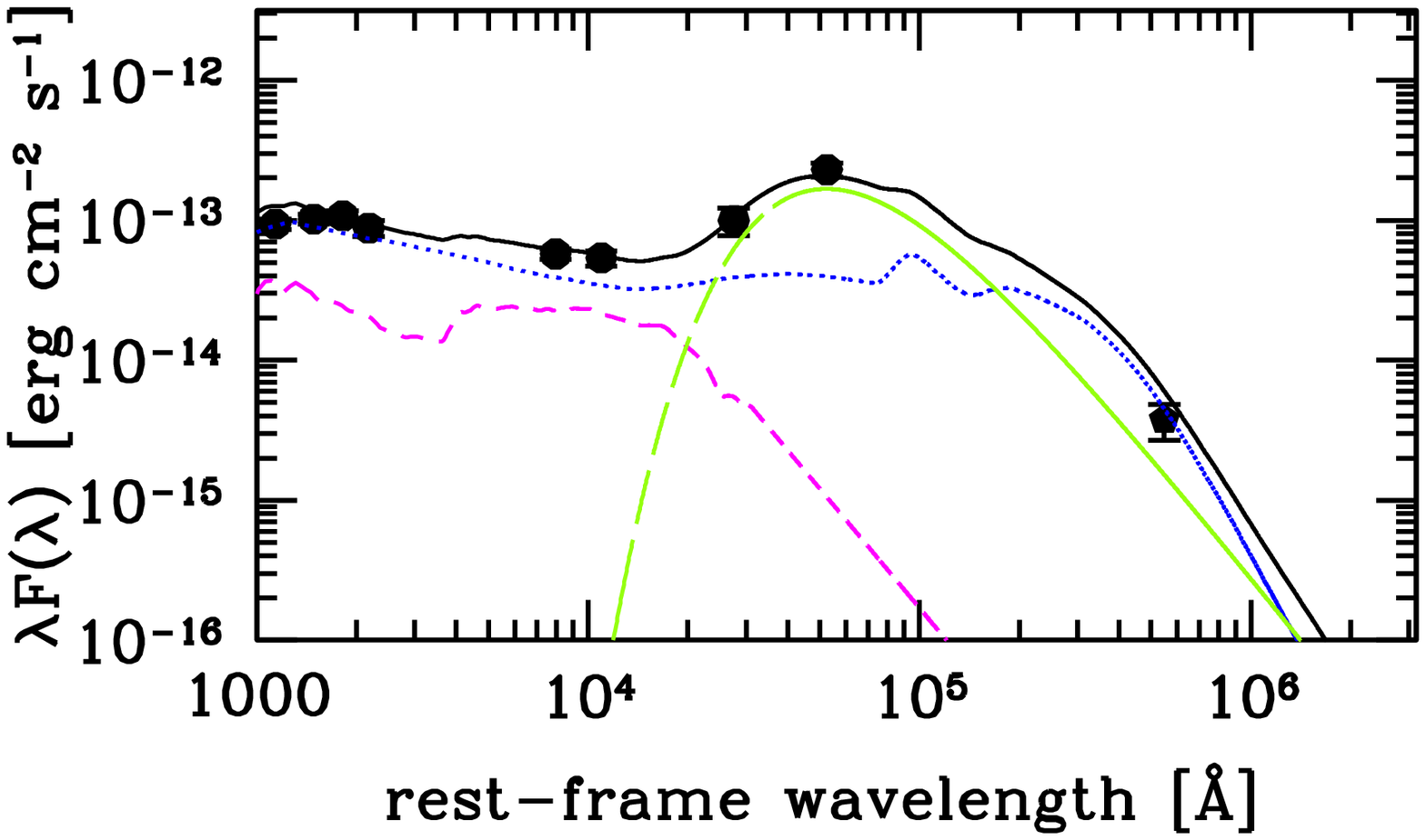}\\
\includegraphics*[scale=0.34]{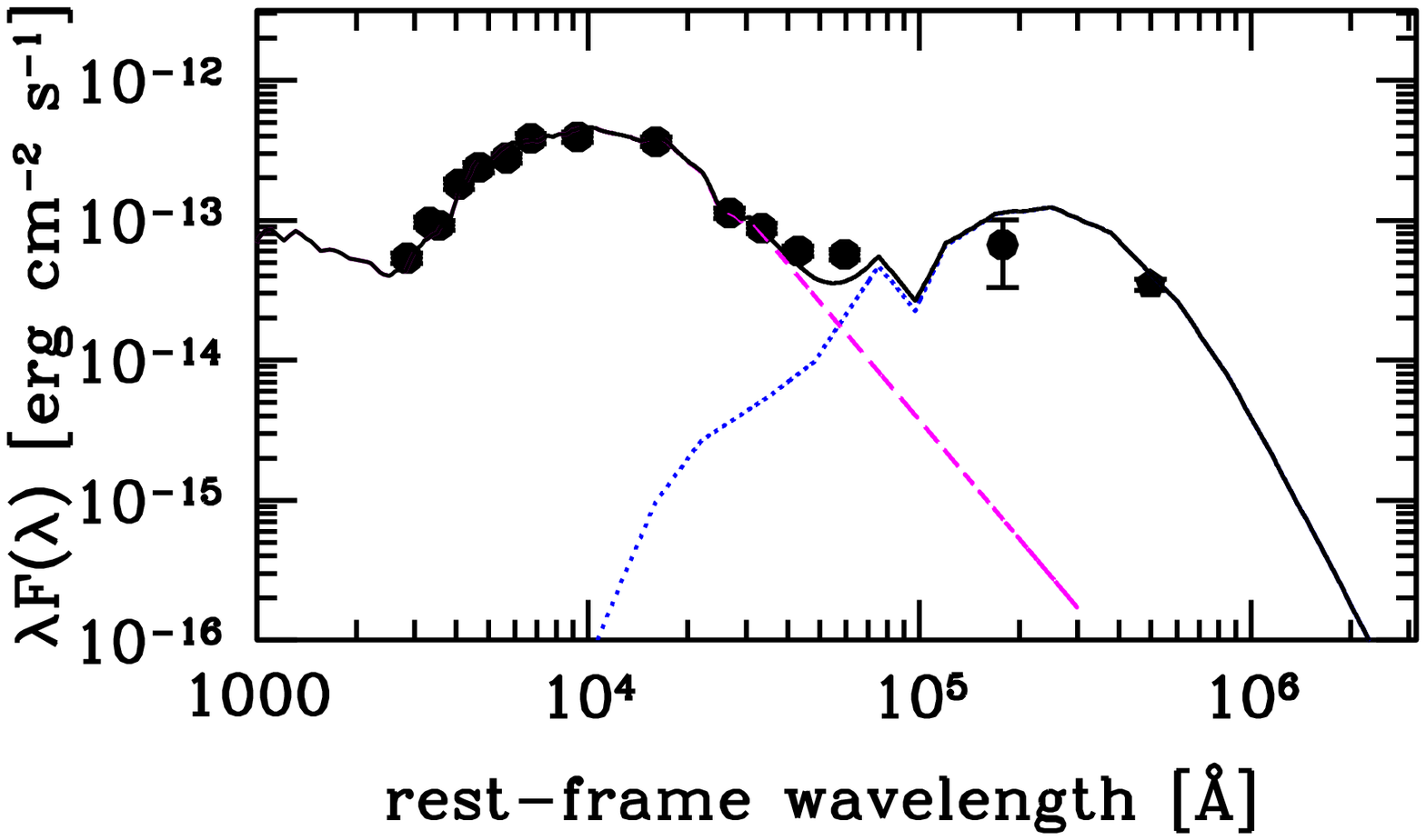}
\caption{Example of SED fitting results for one type 2 source from the SWIFT sample (left top panel), one type 1 from the X-WISSH sample (right top panel), one type 2 from the ASCA sample (left middle panel), one type 1 from the XXL sample (right middle panel) and one type 2 from the COSMOS sample (bottom panel). The AGN, galaxy, cold dust and excess emission components are shown in different colours and line types (as stated in the legend of the first figure on the upper left panel). Black circles are the photometric points of the source. The virtual AGN pivotal photometric point at 50 $\mu$m, derived from the hard X-ray data, is shown as black pentagon (see text for details).}
\label{fig:SED}
\end{figure*}

In D17 the fit was performed with three emission components: 
\begin{enumerate}
    \item the accretion disk plus torus emission \citep[using a combination of models by \citealt{Feltre2012} and][]{Stalevski2016};
    \item the cold dust emission connected to the galactic star formation activity (modelled as a modified blackbody);
    \item an additional component in the MIR, representing warmer dust in the vicinity of the nucleus (described by a simple blackbody).
\end{enumerate}

For the present work we used an improved version of the code described in D17 in order to take also into account, using the photometric information in the X-ray band, sources with low contribution from the nuclear engine. At this aim: (a) we added an additional component, i.e. the UV-optical stellar emission from the host galaxy; (b) the optical-to-MIR AGN SED has been modelled starting from the hard X-ray normalized templates by \citet{Silva2004}.  \\

(a) The galaxy emission component was not included in D17 for the study of type 1 WISSH AGN because for such bright unobscured sources the UV galactic light is completely out-shone by the AGN emission. In this work at variance we added a galaxy component to the three emission components already used in D17. The galaxy emission is produced by the integrated light of diverse stellar populations characterized by distinct Star Formation Histories (SFHs). We generated a library of synthetic spectra, using the stellar population synthesis models of \citet{bc}. We assumed a \citet{Chabrier2003} initial mass function, building 10 declining SFHs as a function of the e-folding time and the age of the galaxy \citep[see e.g.,][]{Bongiorno2012}. For all the galactic templates we took into account the effect of dust extinction inside the galaxy, choosing as reddening curve the Calzetti's law \citep{Calzetti2000}. The final library of galaxy templates used consists of about 900 templates with extinction in the range $0 \leq \rm E(B-V) \leq 0.5$. \\

(b) In type 1 unobscured sources, the AGN component dominates the optical and IR bands compared to the host galaxy emission; on the contrary, in type 2 obscured AGN, the SED is characterized by an optical continuum dominated by the host galaxy emission, while the AGN component is mainly relevant at IR wavelengths. To model the emission from type 2 AGN we adopted the AGN SEDs by \citet{Silva2004} which are derived from a sample of 33 Seyfert galaxies properly corrected for any galaxy nuclear contribution, and available in four intervals of X-ray absorption: $N_{H} < 10^{22} cm^{-2}$ for Seyfert 1 (Sy1), $ 10^{22} < N_{H} < 10^{23} cm^{-2}$, $ 10^{23} < N_{H} < 10^{24} cm^{-2}$ and $ N_{H} > 10^{24} cm^{-2}$ for Seyfert 2 sources (Sy2). Being hard X-ray normalized SEDs, they are able to provide the AGN optical-to-MIR component, once good quality hard X-ray photometric data \citep[corrected for absorption, see][]{Riccic2017}, as in our case, are available. As shown by \citet{Silva2004}, all the AGN X-ray normalized SEDs can be considered quite identical above 40$\mu m$ (see Figure \ref{fig:agn}), with a spread of about 0.1 dex. Assuming an average value of the SEDs at 50$\mu m$, it was therefore possible to convert the hard X-ray photometry into a virtual pivotal photometric, to be fitted only by the AGN templates (see Figure \ref{fig:SED}). 
Aware of this assumption, to overcome the limitation of having only four AGN templates, and in the meantime to better reproduce the spectral emission of either very luminous type 1 or very low luminous type 2 AGN, we enriched the library with the models by \citet{Stalevski2016} used in D17.\\
Summarizing, our emission model ($f_{model}$) includes four emission components: 
\begin{equation}
f_{model} = A  f_{AGN} + B  f_{GAL} + C  f_{CD} + D  f_{ME},
\end{equation}

\noindent
where f$_{AGN}$ represents the emission coming from the active nucleus, both direct and reprocessed by the dusty torus, f$_{GAL}$ is the emission from the stars, f$_{CD}$ accounts for the FIR emission due to the reprocessed flux by cold dust and is modelled as a modified blackbody,  f$_{ME}$ is any eventual excess MIR component, modelled as a simple blackbody. Finally, A, B, C and D are the relative normalizations. The fitting procedure, through a $\chi^2$ minimization, allows to determine the combination of templates that best describes the observed SED and their relative contributions.

\subsection{SED results}
\label{sec:resul}
Based on the final SED best-fit models, the intrinsic AGN bolometric luminosity of the type 1 AGN sources has been derived by integrating the AGN emission component in the range 20 $\AA$ - 1 $\mu$m. Same as L12, we neglected the IR emission in order not to count twice the UV/optical photons reprocessed by the dust \citep[for a different approach see][who instead included this component in the estimation of the total emitted SEDs due to the AGN activity]{Hopkins2007}.  
It is worth noting that the values of the bolometric luminosity published in D17 for the 9 X-WISSH sources having \textit{Herschel} coverage (and in common with this work) are almost identical (within few percents) to those obtained using our modified version of the SED-fitting code, i.e., with the inclusion of the X-ray photometric information.\\
Type 2 AGN emission in the UV/optical bands is known to be more affected by the presence of the host galaxy than type 1 AGN. In this case, to compute the bolometric luminosity we followed the approach by \citet{Pozzi2007}, by integrating the AGN template in the range 1 $\mu$m - 1000 $\mu$m and simply re-scaling the result for a factor 1.9 \citep[see also][]{Vasudevan2010}. If instead the X-ray luminosity is included (as done for the XMM-COSMOS sources by L12), this factor reduces to $\sim 1.7$ \citep[][]{Pozzi2007}. This correction accounts for both the geometry of the torus and its orientation. It is indeed related to the covering factor f (which represents the
fraction of the primary optical-UV radiation intercepted by the
torus) and to the anisotropy of the
IR emission, function of the viewing angle. We checked that, applying this method to the type 1 sources, the resulting bolometric luminosities are consistent (within $8 \%$) with the ones obtained by the direct integration of the UV-to-NIR AGN emission. \\ 
We verified that the SED-fitting method described above provides results similar to the ones obtained by L12 for the COSMOS sample. On a random sample of 10 type 1 and 10 type 2 COSMOS AGN, the bolometric luminosities estimated using our SED-fitting are in agreement with L12 within 0.04 dex. Such a small discrepancy was indeed expected, being the bolometric luminosity simply the area below the AGN emission template, and therefore slightly dependent on the chosen combination of models for the total emission.\\
The values of the derived bolometric luminosity for the SWIFT, X-WISSH, ASCA and XXL samples are reported in Table \ref{tab:tabella}, Table \ref{tab:wissh}, Table \ref{tab:asca} and Table \ref{tab:xxl} in Appendix A. \\

\section{Bolometric correction as a function of luminosity}
\label{sec:kbols}

\begin{figure*}
\centering
\includegraphics[scale=0.70]{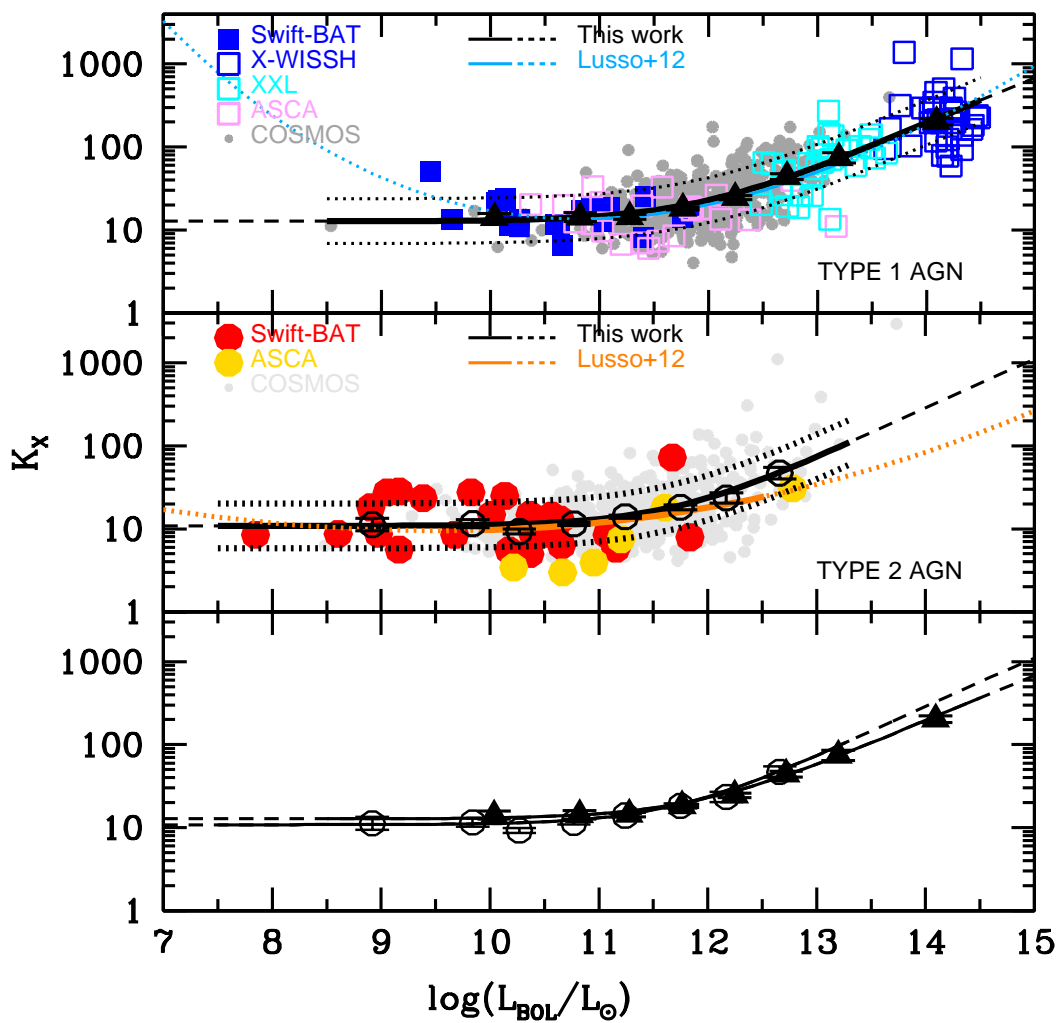}
\caption{Hard X-ray bolometric correction in the 2-10 keV band as a function of the bolometric luminosity for type 1 (upper panel) and type 2 (central panel) AGN.  Symbols are as in legend. Black filled triangles and open circles show the average values for type 1 and type 2 sources respectively (directly compared in the lower panel), in bins of bolometric luminosity. The black solid and dashed lines show our best-fit relations and their extrapolations according to Equation \ref{eq:rela}.}
\label{fig:type1and2}
\end{figure*}

\begin{figure}
    \centering
    \includegraphics[scale=0.40]{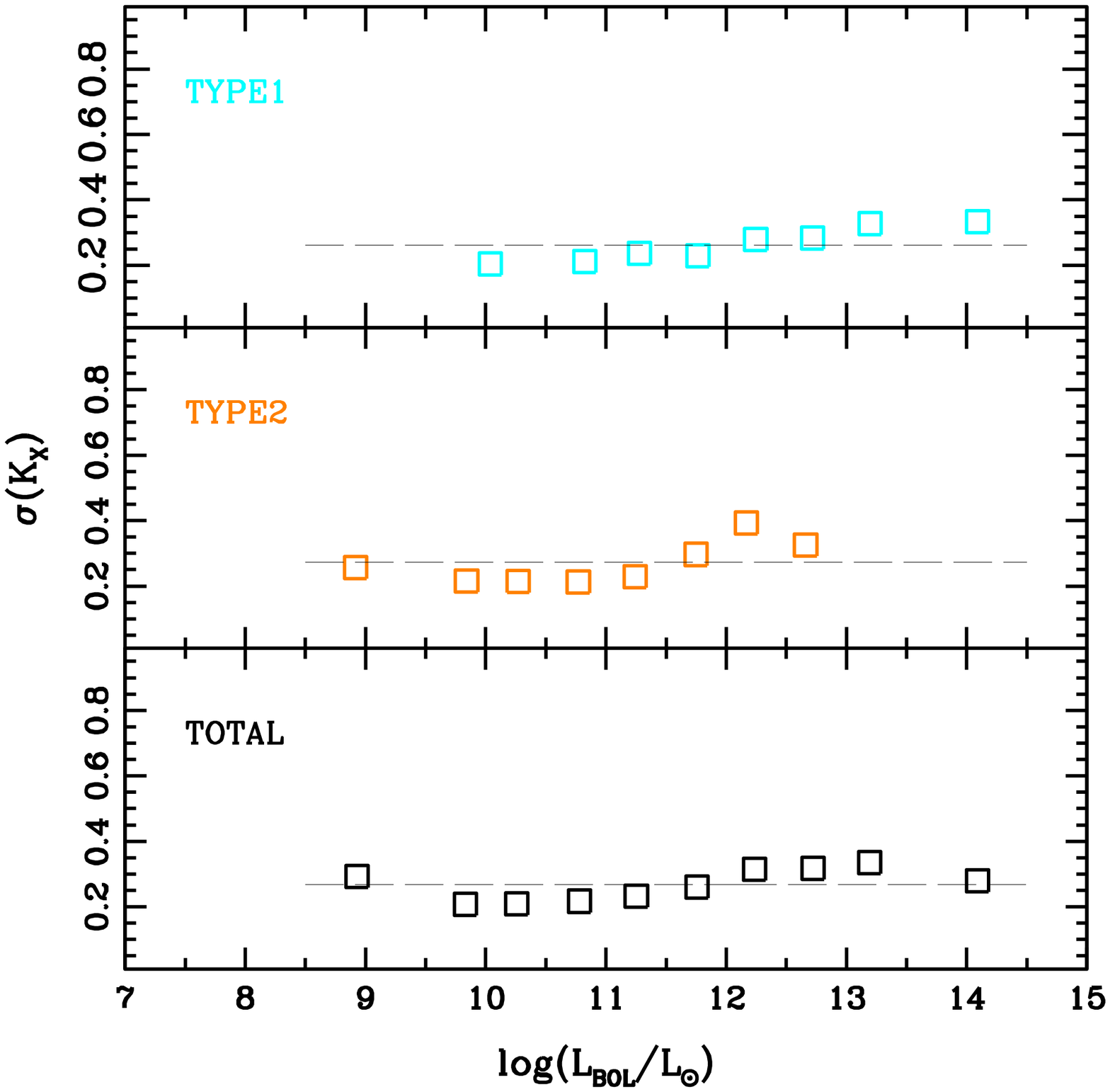}
    \caption{Intrinsic spread in bins of bolometric luminosity for the X-bolometric correction of type 1, type 2 and the whole AGN sample. Dashed black lines show the average value.}
    \label{fig:spread}
    
\end{figure}

The bolometric correction is by definition the ratio between the AGN bolometric luminosity and the luminosity computed in a specific band. It is a quantity of primary importance in astronomy, because it allows us to obtain an estimate of the AGN bolometric luminosity when a complete SED is not available. 

\subsection{Hard X-ray bolometric correction}
\label{sec:kbolx}

\begin{figure*}
\includegraphics[scale=0.80]{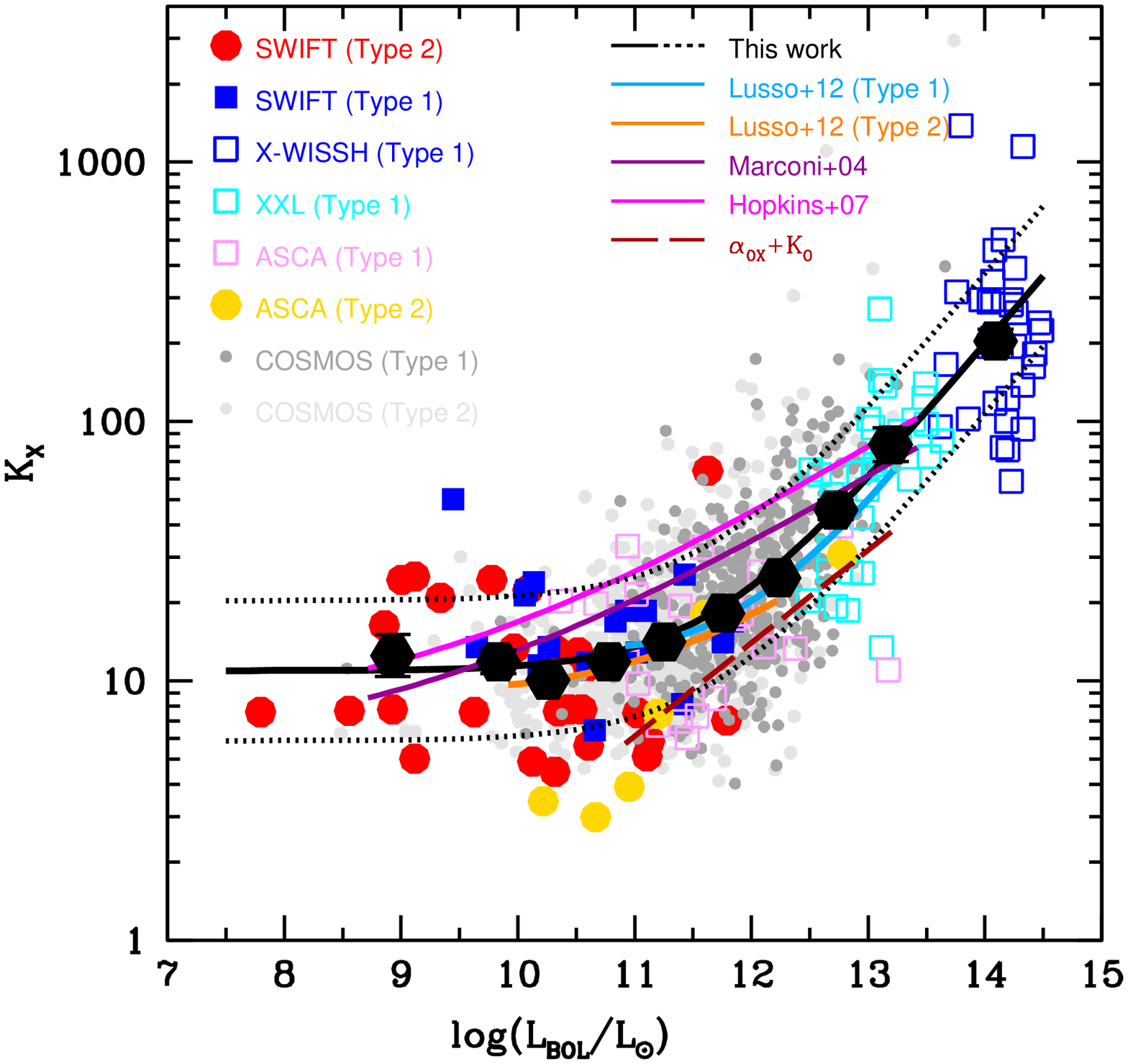}
\caption{General hard X-ray bolometric correction in the 2-10 keV band as a function of the bolometric correction for both type 1 and type 2 AGN. Symbols as in Figure \ref{fig:type1and2}. Black hexagons are the average bolometric correction values, in bins of bolometric luminosity. The black solid line is our best fit solution; the brown solid line is the analytical prediction obtained by assuming the relation between the X-ray luminosity and the optical luminosity by \citet{Lusso2016} and the optical bolometric correction by L12. See text and Table \ref{tab:fit} for details.}
\label{fig:total}
\end{figure*}

As a first step, we have studied the dependence of the AGN hard X-ray bolometric correction on the bolometric luminosity, i.e. $K_{X}(L_{BOL})$. In Figure \ref{fig:type1and2}, we show $K_{X}$ as a function of $L_{BOL}$ for all type 1 sources (upper panel) and type 2 (central panel) sources. Black filled triangles and open circles represent the average of the bolometric corrections computed in bolometric luminosity (not overlapping) bins, for type 1 and type 2 AGN respectively. The light blue continuous line in the upper panel shows the relation found by L12 for their type 1 AGN from the COSMOS sample, while the orange continuous line in the central panel shows the relation found by L12 for their type 2 COSMOS sample; the dotted lines mark the extrapolation of these relations out of the range covered by the L12 data. As clearly visible, the X-WISSH sample and the SWIFT sample cover a range in luminosity never studied before. At high luminosities, as already pointed out in M17, the X-WISSH data fairly follow the extrapolation of the L12 relation at high luminosities, while at low luminosities the SWIFT sample data favour a flatter or constant relation at log($L_{BOL}$/$L_{\odot}$) < 11.\\
We fitted the type 1 and type 2 samples separately, using a least-square method, with the following relation:

\begin{equation}
K_{X}(L_{BOL}) = a \,  \left [1+ \left (\frac{log(L_{BOL}/L_{\odot})}{b} \right )^c \right ] .
\label{eq:rela}
\end{equation}

The best-fit relations for type 1 and type 2 AGN are shown in Figure \ref{fig:type1and2} as a black continuous line and the best-fit parameters, along with the intrinsic average spread of the data ($\simeq 0.27$ dex), are listed in Table \ref{tab:fit}. The values of the spread for both the type 1 and the type 2 samples are quite similar and in good agreement with those found by L12 ($\sim$0.25 for both populations). As it is shown in Figure \ref{fig:spread}, the spread is almost constant throughout the whole luminosity range probed by the data. \\
At low luminosities (in the range 7.5 < log($L_{BOL}$/$L_{\odot}$) < 11) the bolometric correction has a constant value, while it increases at log($L_{BOL}$/$L_{\odot}$) > 11: for type 1 sources, our solution is very similar to the relation found by L12 for their type 1 AGN; on the contrary, for type 2 sources, at log($L_{BOL}$/$L_{\odot}$) > 12.5 it is systematically higher than the one by L12. This might be due to the new functional form used, which allows a better representation of the data sample, and to the fact that in L12 the fitting procedure has been carried out on binned data while we performed the fits considering the whole data-set. 

\noindent

\begin{table}
    \caption{Best-fit parameters for the hard X-ray bolometric correction as a function of either the bolometric luminosity (Equation \ref{eq:rela}) or the hard X-ray luminosity (Equation \ref{eq:rela2}). The luminosities are expressed in solar units.}
    \begin{tabular}{lcccc}
    & a & b & c & spread \\
    & & & & [dex] \\
\hline    
\multicolumn{5}{c}{$K_{X}(L_{BOL})$} \\[2ex]
    Type 1 & 12.76$\pm$0.13 & 12.15$\pm$0.01 & 18.78$\pm$0.14 & 0.26 \\
    Type 2 & 10.85$\pm$0.08 & 11.90$\pm$0.01 & 19.93$\pm$0.29 & 0.27 \\
    General & 10.96$\pm$0.06 & 11.93$\pm$0.01 & 17.79$\pm$0.10 & 0.27 \\
\hline
\multicolumn{5}{c}{$K_{X}(L_X)$ } \\[2ex]
    General & 15.33$\pm$0.06 & 11.48$\pm$0.01 & 16.20$\pm$0.16 & 0.37 \\
\hline

    \end{tabular}
   
    \label{tab:fit}
\end{table}

As it is possible to see in the lower panel of Figure \ref{fig:type1and2}, where the average values of $K_{X}(L_{BOL})$ of type 2 sources (open black circles) are compared with those for type 1 sources (filled black squares), type 1 and type 2 AGN seem to share the same, general, bolometric correction relation. Moreover, as already mentioned above, the two samples have almost the same value of the spread (0.27 dex). We thus investigated whether a general bolometric correction relation, for both type 1 and type 2 sources, was consistent with the data. In  Figure \ref{fig:total} we show the $K_{X}$ as a function of $L_{BOL}$ for the whole sample of type 1 and type 2 AGN. Black filled hexagons represent the average bolometric corrections computed in bolometric luminosity bins for the whole sample. Our best fit relation is shown as a black line, while the two black dotted lines correspond to the 1 $\sigma$ spread of the sample. We found that this general $K_{X}(L_{BOL})$ relation can be considered statistically representative of the whole AGN population. Indeed, it is not rejected by the $\chi^2$ test if applied on the type 1 and type 2 samples separately (P(>$\chi^2$)=0.45 in both cases).\footnote{Since the values of the spread are very similar for the type 1, type 2 sources and whole sample, an average spread of $\sim $ 0.27 dex has been assumed (see Figure \ref{fig:spread}) in computing the $\chi^2$ test.} This result was already pointed out by L12, who argued that the bolometric correction relation for the type 2 sources seems to be the natural extension at lower luminosities of the one for the type 1 sources. Thanks to the inclusion of the low and high luminosity samples used in this work, we have then been able to derive a universal bolometric correction relation valid for the entire X-ray selected AGN population and over the largest (7 dex) luminosity range ever probed.
As already mentioned in Section \ref{sec:sample}, the X-WISSH sample is the only one not being X-ray selected and this might introduce a bias in the resulting bolometric correction. To test if this could cause a significant difference in the final result, we performed a new fit excluding the X-WISSH sources from the total sample, thus minimizing possible biases arising from different selection methods. The difference between the results is negligible, i.e. on average 0.01 dex at the highest luminosities. Therefore, the inclusion of the X-WISSH sources allowed to reduce the uncertainties at the largest luminosities (log$(L_{BOL}/L_\odot)>$13.5) and to extend the validity of the bolometric correction in this extreme regime. 

In Figure \ref{fig:total} the bolometric corrections by L12, for their COSMOS samples of type 1 and type 2 AGN, are shown as cyan and orange lines, respectively. Our result is in good agreement with these relations, if compared within the luminosity range of the COSMOS data used by L12 (11 < log($L_{BOL}$/$L_{\odot}$) < 13). On the contrary, as already found by L12, at log($L_{BOL}$/$L_{\odot}$) $\sim$ 11.5 the \citet[][purple line]{Marconi2004} and \citet[][magenta line]{Hopkins2007} bolometric corrections are about 0.20-0.25 dex above ours. While in the present work the bolometric corrections are directly computed for each object, in \citet{Marconi2004} they have been obtained assuming an $\alpha_{OX}$ relation which was certainly less accurate than today; this might justify the discrepancy (although surprisingly small) with our relation. Concerning \citet{Hopkins2007}, it is worth noticing that they included the IR contribution in the estimate of the bolometric luminosity, being interested in the characterization of the observed SED of the AGN.

The majority of the studies focused on BH accretion and demography (as mentioned in the Introduction) have used the bolometric correction as a function of the luminosity in a specific band. Although it is possible to numerically convert the $K_{X}(L_{BOL})$ into a relation in which the bolometric correction depends on the X-ray luminosity, we directly fitted this relation using the following Equation:

\begin{equation}
K_{X} (L_X) = a \,  \left [1+ \left (\frac{log(L_X/L_{\odot})}{b} \right )^c \right ] .
\label{eq:rela2}
\end{equation}

The parameters and the relative spread are shown in Table \ref{tab:fit}. As discussed in section 2, the 46 AGN of the SWIFT sample are affected by a small bias in favour of sources having, on average, an X/MIR ratio about 0.1 dex larger than the parent  sample.
We have verified that this bias is negligible: the fits differ by only 0.025 dex  once the $L_{BOL}$ values of the SWIFT AGN are reduced by 0.1 dex. 

The $K_{X}(L_X)$ relation provides on average the same X bolometric correction as the $K_{X}(L_{BOL})$ one, with differences within 0.2 dex in the range $8<log(L_{BOL}$/$L_{\odot})<13$ (see also the discussion on the resulting bolometric luminosity function in Section \ref{sec:fdilumin}). However, given the higher value of the spread ($ \sim 0.37$) of this relation, we strongly encourage the reader to adopt the bolometric correction relation as a function of the bolometric luminosity, i.e. Equation \ref{eq:rela}. \\

\subsection{Optical bolometric correction}
\label{sec:kbolo}

\begin{figure}
\includegraphics[scale=0.45]{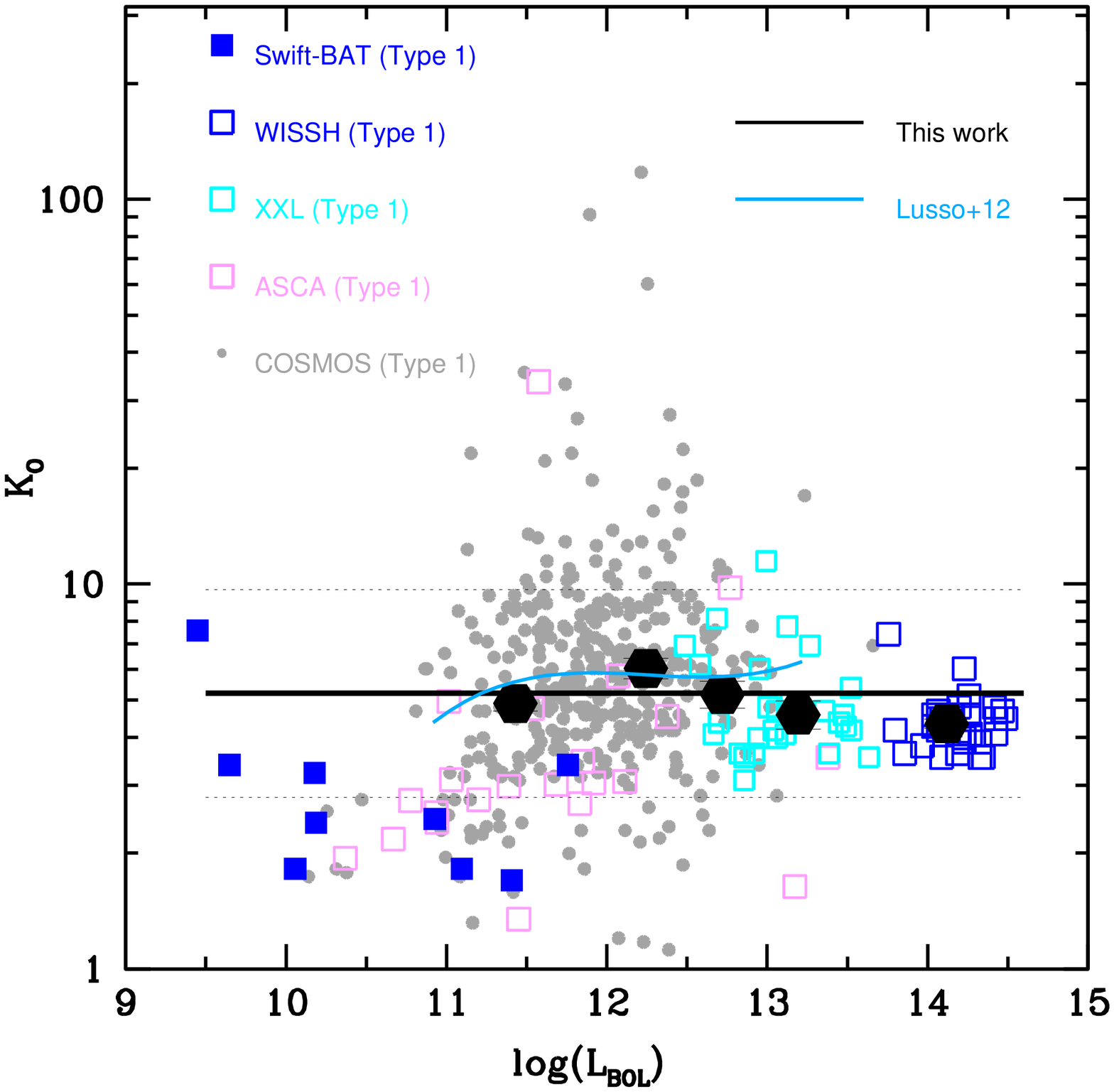}
\caption{Optical bolometric correction computed at 4400 $\AA$ for type 1 sources. The black solid line is our best-fit solution. See text and Table \ref{tab:fito} for details.}
\label{fig:opt}
\end{figure}

In addition to the hard X-ray bolometric correction we also studied the dependence of the optical (specifically in the B-band, at 4400 \,$\AA$) bolometric correction on the bolometric luminosity. We measure the optical luminosity from the photometric point at 4400 $\AA$ rather than from the best AGN model provided by the SED-fitting procedure. In this way, the optical and bolometric luminosities are independent from each other and no artificial relation between them is introduced.
Considering that for type 2 obscured AGN the optical emission is mostly dominated by the host galaxy, we included in this specific analysis only type 1 sources of each sample.\\
In Figure \ref{fig:opt} we show the $K_{O}$ as a function of $L_{BOL}$. Black hexagons represent the average of the bolometric correction in bins of bolometric luminosity. As already found by L12 (whose relation is reported for comparison as a light blue line), the optical bolometric correction does not significantly depend on the bolometric luminosity. Indeed, the Pearson linear correlation test gives a coefficient R equal to $\sim 0.069$, corresponding to a probability of 14 $\%$ to be linearly related. We have then fitted the data with a constant value, using the relation:

\begin{equation}
    K_{O} (L_{BOL}) = k.
\label{eq:opt}
\end{equation}

\noindent
Our best fit relation, $K_{O} (L_{BOL})$ $\sim 5.13$, is shown as a black solid line, while the 1 $\sigma$ spread of the sample ($\sim 0.26$ dex) is marked with the two dotted lines (see Table \ref{tab:fito}).
The values of the bolometric correction at the lowest bolometric luminosities ($log(L_{BOL}/L_{\odot})$ < 11) are about a factor of 2 lower than the rest of the sample. As suggested by L12, this behaviour might be ascribed either to a simple statistical fluctuation, due to the small number of low luminosity sources, or to the fact that in this range the B-band luminosities might be influenced by the contribution of the host galaxy. However, the inclusion or not of these objects does not significantly affect the average value of the optical bolometric correction.\\

We also studied the dependence of the optical bolometric correction on the optical luminosity, $K_{O}(L_O)$. We have fitted the data with a constant value, using the relation:

\begin{equation}
    K_{O} (L_{O}) = k,
\label{eq:opt}
\end{equation}

where $k=5.18 \pm 0.13$, with a spread of $\sim$ 0.27 dex (see Table \ref{tab:fito}). Having very similar intrinsic spread, these two relations could be used without distinction. \\

\begin{table}
\centering
\caption{Best-fit parameters for the optical bolometric correction as a function of either the bolometric luminosity (Equation \ref{eq:opt}) or the optical luminosity.}
\begin{tabular}{llc}
    & k & spread \\
    & & [dex]\\
\hline
    K$_{O}$(L$_{BOL}$) & 5.13$\pm$0.10 & 0.26 \\
    K$_{O}$(L$_{O}$) & 5.18$\pm$0.12 & 0.27 \\
\hline

    \end{tabular}
    \label{tab:fito}
\end{table}

\section{Bolometric correction as a function of redshift}
\label{sec:kbolz}

\begin{figure*}
\centering
\includegraphics[scale=0.50]{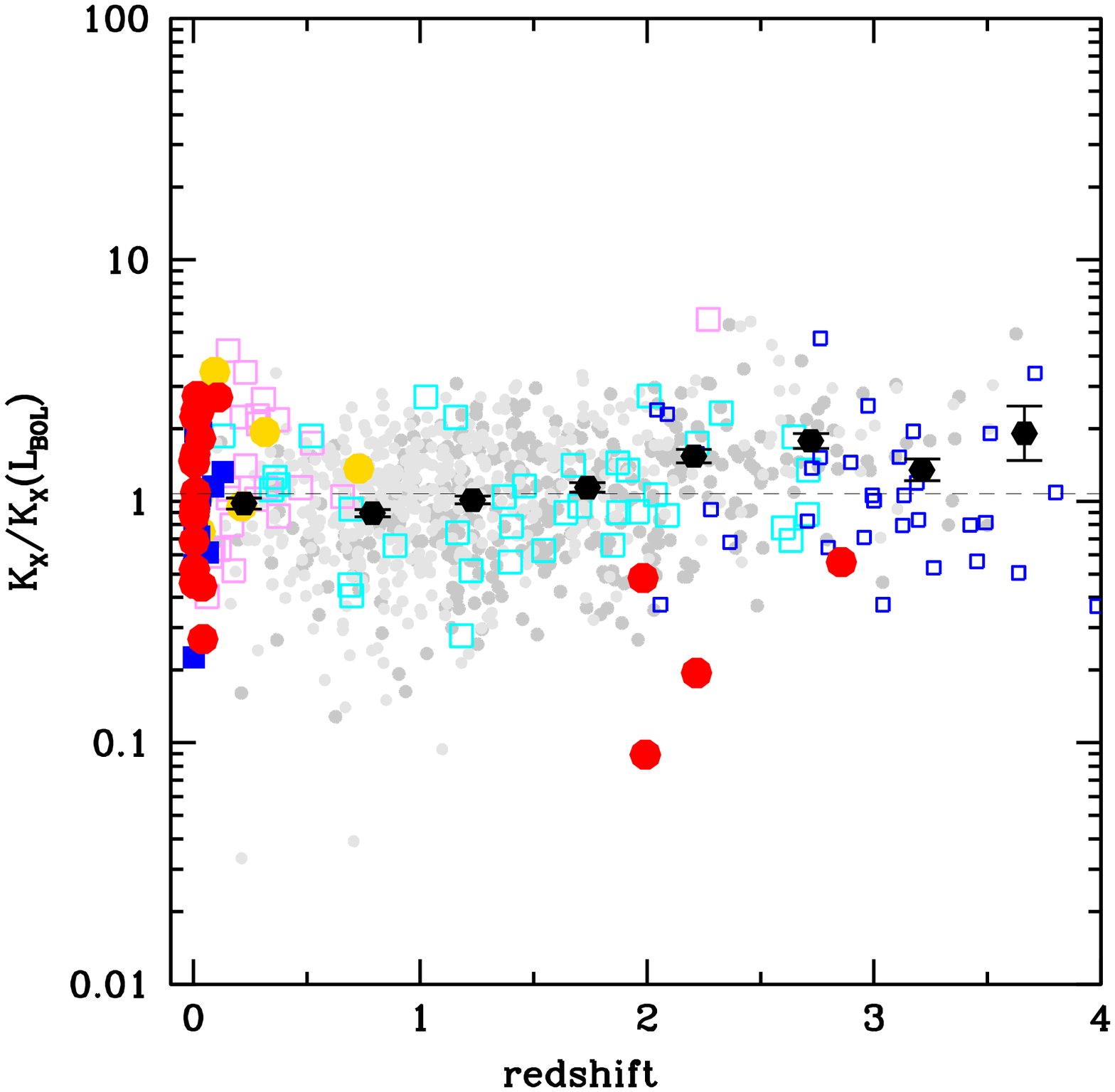}\\
\includegraphics[scale=0.50]{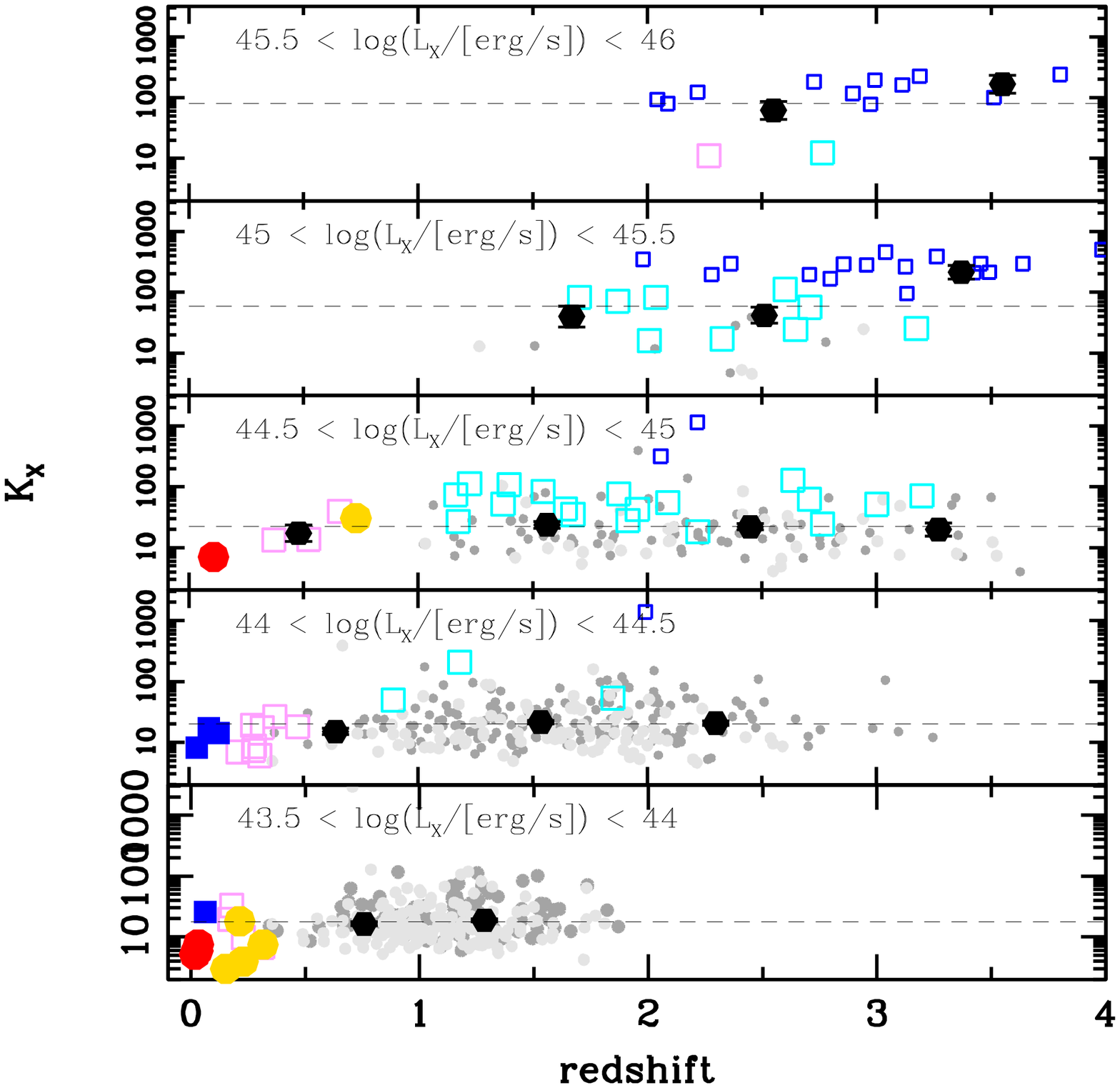}
\caption{Hard X-ray bolometric correction as a function of redshift. In the top panel, the \textit{normalized} bolometric correction has been obtained by dividing $K_{X}$ for the corresponding bolometric correction given by Equation \ref{eq:rela}. Bottom panel shows $K_{X}$ as a function of redshift in five ranges of hard X-ray luminosity. Black hexagons are the average values of $K_{X}$ computed in bins of redshift and the dashed line is the average $K_{X}$ of the whole sample within the entire redshift range. Symbols as in Figure \ref{fig:total}.}
\label{fig:z}
\end{figure*}

We investigated the redshift dependence of the hard X-ray bolometric correction relation. To this end, we computed the normalized bolometric correction by dividing the X-ray bolometric correction of each source by the average value as derived by our best-fit relation (Equation \ref{eq:rela}). The normalized values (black hexagons) and the best-fit (dashed line) are shown in Figure \ref{fig:z} (top panel), and suggest no trend with redshift. As a second test, we divided the whole sample in five ranges of hard X-ray luminosity (bottom panel). Again, black hexagons are the average values of bolometric correction in bins of redshift, while the black dashed lines represent the average value of the bolometric correction of the sample in each range of $L_{X}$. Also in this case we find that in none of the analyzed luminosity ranges there is a hint of dependence on redshift. 

\section{Bolometric correction as a function of Eddington ratio and BH mass}
\label{sec:ledd}

\begin{figure*}
\centering
\includegraphics[scale=0.80]{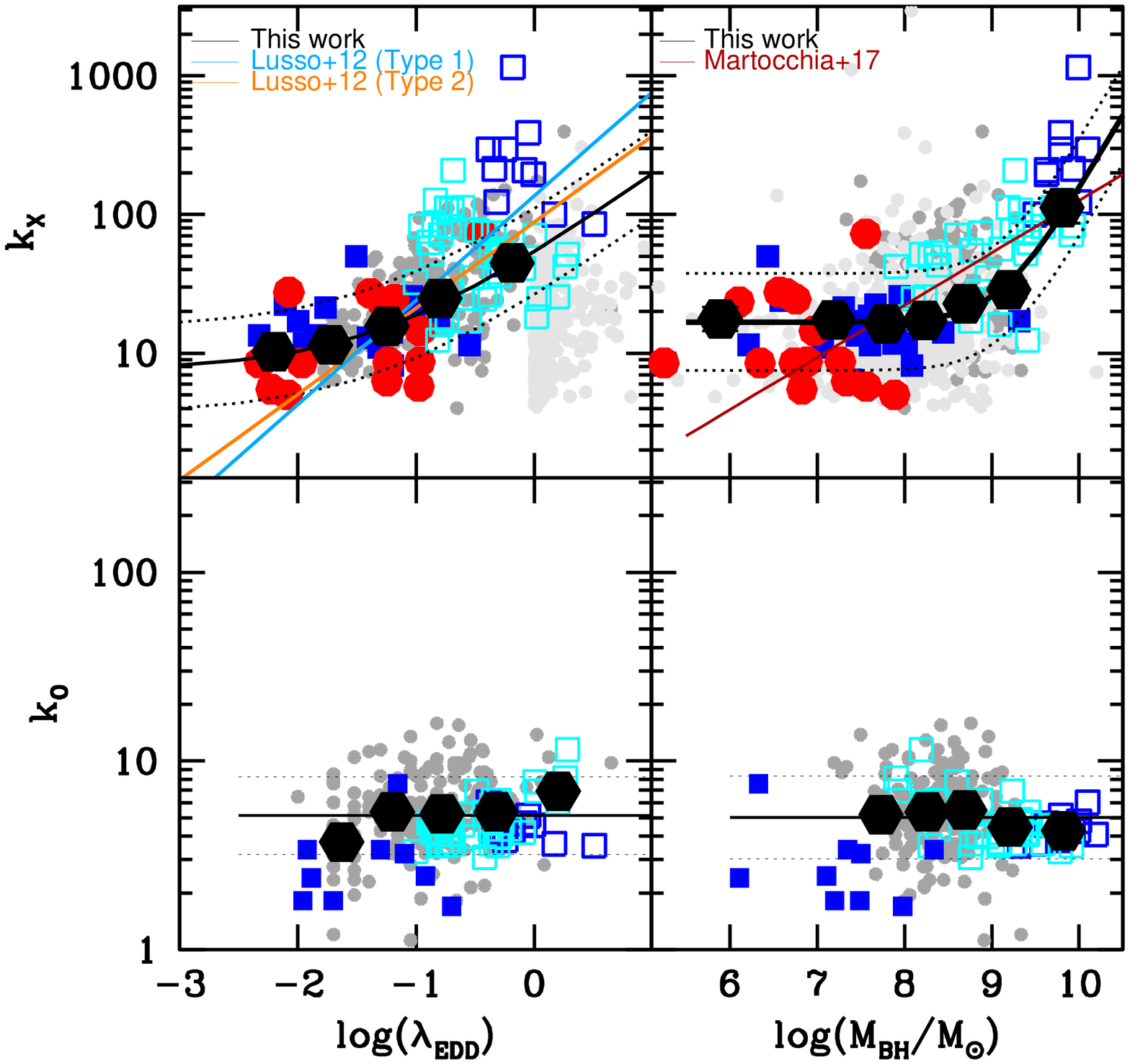} \\
\caption{Upper panels: hard X-ray bolometric correction as a function of the Eddington ratio (left) and of the BH mass (right). Lower panels: optical bolometric correction as a function of the Eddington ratio (left) and of the BH mass (right). Black solid lines show our best fit relations. Symbols are the same as in Figure \ref{fig:total}.}
\label{fig:ledd}
\end{figure*}

Many authors have found a correlation (although with a huge scatter) between the bolometric correction and either the Eddington ratio ($\lambda_{EDD} $ = $L_{bol}$/$L_{EDD}$) or the BH mass \citep[e.g.,][]{Vasudevan2007,Vasudevan2009,Lusso2010}. We studied these dependencies using the same samples presented above, but limited only to the 745 objects for which the $M_{BH}$ has been measured: 21 type 1 and 15 type 2 sources by \citet{Onori2} for the SWIFT/BAT sample (which $M_{BH}$ has been derived from reverberation mapping techniques and deep NIR spectroscopy, see Section \ref{sec:sample}); 11 WISSH type 1 sources \citep[with H$\beta$-based virial masses provided in][]{Vietri2018}; 661  COSMOS AGN by L12 (with masses obtained from virial estimators for the type 1 AGN and from $M_{BH}$ - $M_{*}$ relations plus morphological information for the type 2 sub-sample); and the 37 XXL sources (which have virial masses estimated from $H\beta$, Mg$_{II}$ and C$_{IV}$ broad lines). 
To have measurements as homogeneous as possible, we re-scaled the BH mass values of the SWIFT sample by assuming the same virial factor (f=5.5) as the other samples \citep[rather than f=4.31 as in][]{Ricci2}. Concerning the COSMOS type 2 sample, their BH masses have been derived by L12 using the $M_{BH}-M_{bulge}$ relation by \citet[][]{Haring2004} after correcting for the bulge to total ratio. \footnote{We caution the reader that this last method is different from the one used for all the other samples studied, then potentially this could introduce a bias in the $M_{BH}$ estimates.}
In the upper panels of Figure \ref{fig:ledd} the hard X-ray bolometric correction is shown as a function of the $\lambda_{EDD}$ (left panel) and the $M_{BH}$ (right panel). Black hexagons are the average values of the hard X-ray bolometric correction computed in bins of $\lambda_{EDD}$ and $M_{BH}$. As can be seen from Figure \ref{fig:ledd}, the bolometric correction as a function of the BH mass is only applicable in the range $\lambda_{EDD}>10^{-2}$, as our AGN sample does not cover black holes accreting at lower values of the Eddington ratio. In the upper left panel, the relations published in L12 for their type 1 and type 2 COSMOS AGN are reported as light blue and orange straight lines. As already highlighted in Section \ref{sec:kbolx}, thanks to the new low and high luminosity AGN added in this study, from the data it is now possible to recognize a more complex dependence of the hard X-ray bolometric correction on the $\lambda_{EDD}$, than the one represented by a simple linear relation. We then chose to fit the dependence of $K_{X}$ on $\lambda_{EDD} $ using the relation:

\begin{equation}
K_{X}(\lambda_{EDD}) = a \,  \left [1+ \left (\frac{\lambda_{EDD}}{b} \right )^c \right ] .
\label{eq:relaedd}
\end{equation}

Our best-fit relation (the parameters are provided in Table \ref{tab:fitedd}), is showed as a black solid line. Its average intrinsic spread, $\sim$ 0.31 dex (black dotted lines), is larger than the one previously found for the fitting relation using the bolometric luminosity as the independent variable ($\sim$ 0.27; see Section \ref{sec:kbolx}). \\

The same analysis has been performed for the $K_{X}$ - $M_{BH}$ relation (right panel). Again, we find that a linear relation is not a good representation of the whole data set (see, as an example, the relation found by M17), because it would underestimate $k_{X}$ at both low $\lambda_{EDD}$ and low $M_{BH}$. We then used the following equation to fit the data:

\begin{equation}
K_{X}(M_{BH}) = a \,  \left [1+ \left (\frac{log(M_{BH}/M_{\odot})}{b} \right )^c \right ] .
\label{eq:relambh}
\end{equation}

\begin{table}
    \centering
    \caption{Best-fit parameters (see Equation \ref{eq:rela}) for the hard X-ray bolometric correction relation as a function of the Eddington ratio and the BH mass.
    }
    \begin{tabular}{lcccc}
    & a & b & c & spread \\
    & & & & [dex] \\
\hline
    $\lambda_{EDD}$ & 7.51$\pm$1.34 & 0.05$\pm$0.03 & 0.61$\pm$0.07 & 0.31 \\
    M$_{BH}$ & 16.75$\pm$0.71 & 9.22$\pm$0.08 & 26.14$\pm$3.73 & 0.35 \\
\hline
    \end{tabular}
   
    \label{tab:fitedd}
\end{table}

The best-fit relation (see Table \ref{tab:fitedd}) is showed as a black solid line. Its average intrinsic spread, $\sim$ 0.35 dex (black dotted lines), is larger than both the one previously found for the $K_{X}(\lambda_{EDD})$ and the $K_{X}(L_{BOL})$ relations (0.31 dex and 0.27 dex respectively).\\
In the lower panels of Figure \ref{fig:ledd} we show the optical bolometric correction as a function of the $\lambda_{EDD}$ (left panel) and the $M_{BH}$ (right panel). Black hexagons are the average values computed in bins of Eddington ratios and BH masses. As already found for $K_O(L_{BOL})$ the optical bolometric correction shows a roughly constant average value, $ k \sim$ 5, with a spread of $\sim$ 0.22 dex (see Table \ref{tab:fiteddo}). \\

\begin{table}
\centering
\caption{Best-fit parameters (see Equation \ref{eq:opt}) for the optical bolometric correction relation as a function of the Eddington ratio and the BH mass.}
\begin{tabular}{llc}
    & k & spread \\
    & & [dex]  \\
\hline
$\lambda_{EDD}$ & 5.10$\pm$0.13 & 0.22 \\
    M$_{BH}$ & 5.05$\pm$0.35 & 0.21 \\
\hline
    \end{tabular}
    \label{tab:fiteddo}
\end{table}

\section{The AGN unobscured bolometric luminosity function}
\label{sec:fdilumin}

\begin{figure}
\includegraphics[scale=0.70]{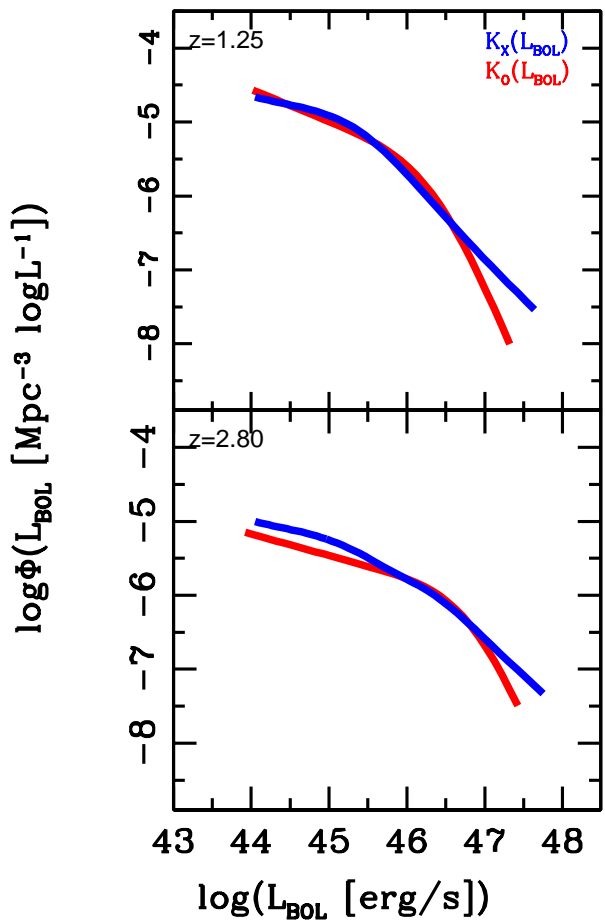} \\

\caption{AGN type 1 bolometric luminosity function computed in the two redshift bins (z=1.25 and z=2.80) adopting the X-ray and optical bolometric corrections as a function of the bolometric luminosity.}
\label{fig:fdilum}
\end{figure}

To verify the consistency of the above computed bolometric corrections, we derived the type 1 AGN bolometric luminosity function (LF) to see if compatible results can be obtained starting from either optically selected or X-ray selected AGN LFs. \\
As shown by \citet{Ricci2017}, the type 1 optical LF is in agreement with the X-ray LF, when only the unabsorbed ($N_H < 10^{21} - 10^{22} $ cm$^{-2}$) sources are considered and the $\alpha_{OX}$ relation as computed by \citet{Lusso2016} is adopted, convolved with an observational scatter of $\sim$0.4~dex. \\
Based on the aforementioned result, if the bolometric corrections derived in this work were consistent, the bolometric LFs computed either with $K_O$ or $K_X$ should produce similar LF in the bolometric domain. We show in Figure \ref{fig:fdilum} the resulting bolometric LFs in two representative redshift bins: $z\sim1.25$ and $\sim2.8$. The optical LF was parametrized with the double-powerlaw as derived by \citet{Palanque2013} for redshift 1.25 and as derived by \citet{Ross2013} at $z\sim2.80$. 
When needed, we converted the g-band optical LF into i-band $z=2$ absolute magnitude adopting Equation 8 of \citet{Ross2013}, then we converted them into 2500~$\AA$ fluxes using the standard methodology outlined in \citet{Richards2006}, and finally converted into 4500~$\AA$ fluxes assuming a power-law SED $L_\nu \propto \nu^{-\alpha}$ with $\alpha=0.5$ \citep[][check footnote \ref{footnote:xx}]{Berk2001}. Then we applied the best-fit optical bolometric corrections derived in Section \ref{sec:kbolo} and convolved with its spread. The resulting curves are shown in Figure \ref{fig:fdilum} as red solid lines.\\
For the X-ray selected LF, we adopted the \citet{Ueda2014} 2-10 keV LF, which was computed in various absorption ranges (i.e. at different 
$N_H$), so in order to match the optically selected LF, only the AGN with $N_H<10^{21}$~cm$^{-1}$ were selected \citep{Ricci2017}. We then converted the X-ray luminosity function into the bolometric LF using the $K_{X}(L_{BOL})$ (blue line in Figure \ref{fig:fdilum}) bolometric correction derived in Section \ref{sec:kbolx}, taking into account its intrinsic spread.\\
The two luminosity functions are in fairly good agreement within 0.2 dex in space density in the range 44<$log(L_{BOL}/[erg/s])$<47.5. \\
A more accurate statistical analysis and precise computation of the AGN bolometric LF, taking into account the spread of the luminosity distributions and selection effects, is beyond the scope of this paper \citep[see e.g. recent results on this topic by][]{Shen2020}. However, the above findings lend further support to the validity of our bolometric corrections.

\section{DISCUSSION}
\label{sec:disc}

The main result of our analysis is that type 1 and type 2 sources share the same bolometric correction in their $\sim$ 4 dex wide overlapping luminosity regime. This allowed us to build a universal bolometric correction valid for about 7 dex in luminosity, where the low luminosity regime is only probed by type 2 AGN and the high luminosity regime by type 1 AGN. The similarity between the bolometric correction relations found for type 1 and type 2 AGN seems to favour the classical AGN unification scenario \citep{Antonucci1993, Urri1995,Antonucci2012}, or at least any model in which the central emission mechanism is the same for both AGN populations. \\

Another interesting result is that the computed bolometric corrections (and then the central emission mechanisms) do not show any significant dependence on redshift up to $z\sim3.5$, as expected if the bolometric corrections are determined by the physics of accretion and emission. Therefore, they can be safely used to study the accretion rate histories of the AGN in the last 11 Gyr. \\

Both the optical and the hard X-ray bolometric corrections follow, separately, the same analytical behaviour whatever the independent variable is chosen. Indeed, while the former remains always nearly constant, the latter shows a significant increase for the most luminous and massive AGN. \\

The ratio between the bolometric luminosity and the X-ray luminosity in AGN can be seen as the ratio between the total energy output dominated by the big blue bump component due to thermal UV emission from the accretion disk and the emission released by the  X-ray corona. 
Accordingly,  the correlation we observe between $K_{X}$ and $L_{BOL}$ supports a scenario in which the more the accretion disk emits, the less the corona radiative power contributes to the total energy output.
This may suggest a change in the properties of the X-ray corona in the most powerful AGN possibly related to a stronger
photon-trapping and advection of X-ray radiation in case of high accretion rates or the presence of UV radiation-driven accretion-disk winds affecting the corona (see M17, \citealt{Nardini2019} and Zappacosta et al. submitted). This behaviour seems to be typical of sources with extreme values of the total bolometric luminosity. Our findings are in agreement with the existence of a correlation between the $\alpha_{OX}$ and the optical luminosity (estimated at 2500 $\AA$), as showed e.g. in the work by \citet{Lusso2016}.
We have then computed the expected $K_{X}$ if the relation between the X-ray luminosity and the optical luminosity by \citet{Lusso2016} (in which $L_X \propto$ $L_{UV}^{0.6}$) and a constant $K_{O}$ $\sim$ 5 (see the next Section for further details) are assumed\footnote{\label{footnote:xx} $L_{O}$ has been converted into $L_{UV}$ =  $L_{2500 \AA}$ by assuming a spectral slope of $\lambda^{-1.54}$ \citep{Berk2001}.}. As shown in Figure \ref{fig:total} as a brown line, the resulting analytical prediction, i.e. $K_{X}$ $\propto$ $L_{BOL}^{0.358}$, is in fair agreement with our best-fit relation, sharing, within the uncertainties, similar slope and normalization. 
This comparison tells us that the bolometric correction relation and the $\alpha_{OX}$ slope are tightly related: the observed correlations are mainly due to the change of the X-ray emission fraction in the bolometric luminosity budget. In this framework, the X-ray bolometric correction is a better measure of this phenomenon, as it takes into account the whole energetic emission and does not suffer from the obscuration which could affect the UV/optical luminosity (especially in type 2 sources).\\

\section{Summary and Conclusions}
\label{sec:concl}

In this paper we combined five AGN samples to characterize a general bolometric correction valid for the entire AGN population. We considered a total of 1009 AGN sources (501 of which are type 1 sources and the remaining 566 type 2) properly selected to cover the widest luminosity range ever sampled (with $41 < log(L_{BOL}/[erg/s]) < 48$), and within a redshift range from $z\sim0$ up to $z\sim4$. For 745 of these sources we have an estimate of the $M_{BH}$.\\ We carried out a dedicated SED-fitting procedure.

Our main results are the following:

\begin{itemize}
    \item Concerning the hard X-ray bolometric correction, $K_{X}$, we confirm that it correlates with the bolometric luminosity $L_{BOL}$ for both the type 1 and the type 2 AGN populations, being fairly constant at $log(L_{BOL}/L_{\odot}) < 11$ while it increases up to an order of magnitude at $log(L_{BOL}/L_{\odot}) \sim 14.5$. The relations found for the type 1 and the type 2 samples are very similar and present the same average intrinsic spread ($\sim$ 0.27 dex). Therefore, it has been possible to derive a universal hard X-ray bolometric correction, valid for type 1 and type 2 AGN.
    
    \item The $K_{X}$ correction as a function of the hard X-ray luminosity gives, on average, the same X bolometric correction to the one in which the $L_{BOL}$ is the independent variable, with differences within 0.2 dex in the range $8<log(L_{BOL}/L_{\odot})<13$.
    
    \item We discussed how the positive trend we observe between the X-ray bolometric correction and the bolometric luminosity may be fitted in the "X-ray weakness" scenario, according to which the X-ray emitting corona tends to be less powerful than the UV/optical emission, with increasing bolometric luminosity.
    
    \item The dependence of the optical (B-band, at 4400 $\AA$) bolometric correction either on the bolometric luminosity or the optical luminosity shows an almost constant (i.e. $K_{O} \sim 5$) behaviour, with average intrinsic spread of $\sim 0.27$ dex.
    
    \item We find that $K_{X}(L_{BOL})$ increases with increasing $\lambda_{EDD}$ and $M_{BH}$, while $K_{O}(L_{BOL})$ is roughly constant with an average value of $\sim 5$. This analysis suggests that the optical and the hard X-ray bolometric corrections follow, separately, the same analytical behaviour, independently of which variable they are compared to.
    
    \item We find no dependence of both the X-ray and the optical bolometric corrections on the redshift up to $z\sim3.5$. 
    
    \item We derive the type 1 AGN bolometric luminosity function by converting the hard X-ray (optical) luminosity function using the hard X-ray (optical) bolometric correction as a function of the bolometric luminosity. We find that the X-ray and optical luminosity functions are in agreement within 0.2 dex in the range $44<log(L_{BOL}/[erg/s])<47.5$. We repeat the same analysis using the bolometric corrections as a function of the hard X-ray and optical luminosities, and retrieving, also in this case, a similar good match.
    
\end{itemize}

We thus provide a universal bolometric correction, statistically representative of the entire X-ray selected AGN population, which spans about 7 decades in luminosity and can be applied up to $z \sim 3.5$ in order to estimate the energetic budget of the AGN and their accretion rate history.  

\begin{acknowledgements}
F. Ricci acknowledges financial support from
FONDECYT Postdoctorado 3180506 and CONICYT project Basal AFB-170002.\\
E. Piconcelli, A. Bongiorno, S. Bianchi  acknowledge financial support from ASI and INAF under the contract 2017-14-H.0 ASI-INAF. \\
C. Vignali acknowledges financial support from ASI-INAF I/037/12/0 and ASI-INAF n.2017-14-H.0.
F. Shankar acknowledges partial support from a Leverhulme Trust Research Fellowship. \\
\end{acknowledgements}



\bibliographystyle{aa}
\bibliography{kbol_duras} % if your bibtex file is called example.bib

%%%%%%%%%%%%%%%%%%%%%%%%%%%%%%%%%%%%%%%%%%%%%%%%%%%%%%%%%%%%%%

\begin{appendix}
\section{Physical properties of the samples}

\begin{table*}
\centering
\begin{threeparttable}
\caption{Properties of the SWIFT type 2 and type 1 sources.} 
\label{tab:tabella} 
%\toprule\toprule
\begin{tabular}{lccccc}
\hline
{Name} & $z$ &  log$L_{X} $ &  log$L_{BOL}$  & log$L_{O} $ & logM$_{BH}$ \\
(1) & (2) & (3) & (4) & (5) & (6)  \\
\hline
\hline

	&		\textbf{Type 2 sources} &	\\[2ex]

IRAS F 05189-2524	    & 0.0426 	& 43.40 &  45.26 & 	-	&	7.45	\\
2MASX J07595347+2323241 & 0.0292 	& 43.25 &  43.95 & 	-	&	7.78	\\
MCG -01-24-12 			& 0.0196 	& 43.24 &  44.18 & 	-	&	7.16	\\
NGC 1275 				& 0.0176 	& 43.98 &  44.74 & 	-	&	7.46	\\
Mrk 348 				& 0.0150 	& 43.44 &  44.24 & 	-	&	7.23	\\
Mrk 1210 				& 0.0135 	& 43.13 &  44.07 & 	-	&	6.78	\\
ESO 234-G -050 			& 0.0088 	& 41.60 &  42.97 & 	-	&	6.00	\\
NGC 2992 				& 0.0077 	& 42.00 &  42.74 & 	-	&	6.72	\\
NGC 7465 				& 0.0065 	& 41.97 &  43.40 & 	-	&	6.54	\\
NGC 5506 				& 0.0062 	& 42.99 &  44.15 & 	-	&	6.86	\\
NGC 1365 				& 0.0055 	& 42.32 &  43.71 & 	-	&	6.65	\\
NGC 1052 				& 0.005 	& 41.62 &  42.55 & 	-	&	6.63	\\
NGC 6221 				& 0.0050 	& 41.20 &  42.64 & 	-	&	6.46	\\
NGC 7314 				& 0.0048 	& 42.33 &  43.26 & 	-	&	6.24	\\
NGC 4395 				& 0.0013 	& 40.50 &  41.43 & 	-	&	5.14	\\

                                                       					\\
PKS 0326-288            & 0.108		& 44.52	&  45.42 & 	-	&	-		\\
NGC3079                 & 0.0372	& 41.30	&  42.75 & 	-	&	-		\\
Mrk417                  & 0.0327	& 43.73	&  44.65 & 	-	&	-		\\
NGC612                  & 0.0298	& 43.94	&  44.76 & 	-	&	-		\\
NGC788                  & 0.0136	& 43.02	&  43.76 & 	-	&	-		\\
NGC3281                 & 0.0107	& 43.12	&  44.23 & 	-	&	-		\\
NGC4388                 & 0.0084	& 43.05	&  43.98 & 	-	&	-		\\
ESO 005-G004            & 0.0062	& 42.78	&  43.95 & 	-	&	-		\\
NGC5643                 & 0.004		& 42.43	&  43.60 & 	-	&	-		\\
NGC4941                 & 0.0037	& 41.25	&  42.18 & 	-	&	-		\\
NGC4138                 & 0.003		& 41.23	&  42.49 & 	-	&	-		\\
         
\hline                                              
	& \textbf{Type 1 sources} &	\\[2ex]                                       
         
Mrk 876 				&  0.129  	& 44.19  &  45.34 &   - 		&  8.54  \\
Mrk 1383				&  0.087	& 44.19	 &	45.42 &   44.89 	&  9.31  \\	
Mrk 771 				&  0.063  	& 43.60  &  45.01 &   - 		&  7.94  \\
Mrk 509 				&  0.034  	& 44.08  &  44.99 &   44.48 	&  8.08  \\
Mrk 817 				&  0.031  	& 43.49  &  44.60 &   44.08 	&  7.90  \\
Mrk 279 				&  0.030  	& 43.41  &  44.68 &   44.28 	&  7.59  \\
Mrk 335 				&  0.026  	& 43.24  &  44.51 &   44.29 	&  7.21  \\
Mrk 590 				&  0.026  	& 42.70  &  43.76 &   - 		&  7.60  \\
Mrk 79 					&  0.022  	& 43.11  &  44.18 &   43.92 	&  8.02  \\
NGC 5548				&  0.017  	& 43.14  &  44.21 &   43.95 	&  7.87  \\
NGC 7469				&  0.016  	& 43.19  &  44.42 &   44.14 	&  7.42  \\
NGC 4748				&  0.015  	& 42.34  &  43.72 &   - 		&  6.58  \\
NGC 4253				&  0.013  	& 42.71  &  43.77 &   - 		&  6.21  \\
NGC 3783				&  0.010  	& 43.43  &  44.24 &   43.37 	&  7.38  \\
NGC 3516				&  0.009  	& 42.72  &  43.85 &   - 		&  7.59  \\
NGC 4593				&  0.009  	& 42.81  &  43.72 &   - 		&  7.06  \\
NGC 6814				&  0.005  	& 42.31  &  43.64 &   - 		&  7.30  \\
NGC 3227				&  0.004	& 42.10	 &	43.23 &   - 		&  7.45  \\
NGC 4151				&  0.003  	& 42.31  &  43.66 &   43.28		&  7.66  \\
NGC 4051				&  0.002  	& 41.33  &  43.03 &   - 		&  6.43  \\

\hline
\end{tabular}

\begin{tablenotes}
\small
\item (1) Source name; (2) redshift \citep{Baumgartner2013}; (3) intrinsic logarithmic hard X-ray luminosity (2-10 keV) in erg/s units from \citet{Riccic2017}; (4) SED-fitting derived logarithmic bolometric luminosity in erg/s units; (5) B-band 4400 $\AA$ AGN logarithmic luminosity in erg/s units from the photometric data, provided only for type I sources, as the contamination from the galaxy is negligible; (6) logarithmic $M_{BH}$ in $M_{\odot}$ from \citet{Onori2} for the type 2 sources and from \citet{Ricci2} for the type 1 sources.
\end{tablenotes}
\end{threeparttable}
\end{table*}

\begin{table*}
\centering
\begin{threeparttable}
\caption{Properties of the X-WISSH sample.} 
\label{tab:wissh}
\begin{tabular}{llcccc}
\hline
{Name} & $z$ &  log$L_{X} $ &  log$L_{BOL}$  & log$L_{O} $ &  logM$_{BH}$ \\
(1) & (2) & (3) & (4) & (5) & (6)  \\

\hline
\hline
J0045+1438	&	1.992	&	44.24	&	47.38 & 46.76 &  -\\
J0209-0005	&	2.856	&	45.16	&	47.62 & 46.99 &  - \\
J0735+2659	&	1.982	&	45.11	&	47.65 & 47.03 &  - \\
J0745+4734	&	3.225 *	&	46.37	&	48.01 & 47.40 &  10.20     \\
J0747+2739	&	4.11 	&	45.43	&	47.44 & 46.88 &  - \\
J0801+5210	&	3.263 *	&	45.25	&	47.84 & 47.13 &  9.79    \\
J0900+4215	&	3.294 *	&	46.00	&	47.93 & 47.38 &  9.32    \\
J0904+1309	&	2.974	&	45.89	&	47.78 & 47.17 &  - \\
J0947+1421	&	3.040	&	45.01	&	47.67 & 47.02 &  - \\
J1014+4300	&	3.126	&	45.43	&	47.85 & 47.24 &  - \\
J1027+3543	&	3.112	&	45.79	&	48.00 & 47.32 &  - \\
J1057+4555	&	4.140	&	45.77	&	47.91 & 47.36 &  - \\
J1106+6400	&	2.220 *	&	45.69	&	47.78 & 47.10 &  10.00    \\
J1110+4831	&	2.957	&	45.36	&	47.81 & 47.21 &  - \\
J1111+1336	&	3.492 *	&	45.36	&	47.69 & 47.06 &  9.93    \\
J1159+1337	&	3.984	&	45.04	&	47.74 & 47.18 &  - \\
J1200+3126	&	2.993	&	45.55	&	47.84 & 47.18 &  - \\
J1201+1206	&	3.512 *	&	45.77	&	47.77 & 47.21 &  9.51    \\
J1215-0034	&	2.707	&	45.33	&	47.62 & 46.96 &  - \\
J1236+6554	&	3.424 *	&	45.33	&	47.65 & 46.98 &  9.63    \\
J1245+0105	&	2.798	&	45.03	&	47.25 & -     &  - \\
J1249-0159	&	3.638	&	45.08	&	47.55 & 46.97 &  - \\
J1250+2631 	&	2.044	&	45.94	&	47.91 & 47.32 &  - \\
J1328+5818  &	3.133	&	45.22	&	47.20 & -     &  - \\
J1333+1649	&	2.089	&	45.83	&	47.73 & 47.07 &  - \\
J1421+4633	&	3.454 *	&	45.18	&	47.65 & 47.09 &  9.79 \\
J1426+6025  &	3.189	&	45.72	&	48.07 & 47.42 &  - \\
J1441+0454	&	2.059	&	44.84	&	47.34 & 46.47 &  - \\
J1513+0855	&	2.897	&	45.60	&	47.67 & 47.12 &  - \\
J1521+5202  &   2.218   &   44.85   &   47.91 & 47.28 &  9.99 \\
J1549+1245	&	2.365 *	&	45.34	&	47.81 & 47.03 &  10.10 \\
J1621-0042	&	3.710	&	46.04	&	47.81 & 47.22 &  - \\
J1639+2824	&	3.801	&	45.67	&	48.05 & 47.38 &  - \\
J1701+6412	&	2.727	&	45.75	&	48.01 & 47.36 &  - \\
J2123-0050	&	2.281 *	&	45.43	&	47.72 & 47.09 &  9.59 \\

\hline

\end{tabular}
\begin{tablenotes}
\small
\item (1) Source name; (2) redshift from SDSS or from \citet{Vietri2018} (marked with an asterisk); (3) intrinsic logarithmic hard X-ray luminosity (2-10 keV) in erg/s units, from \citet{Martocchia2017}; (4) SED-fitting derived logarithmic bolometric luminosity in erg/s units; (5) B-band 4400 $\AA$ AGN logarithmic luminosity in erg/s units, from the photometric data when present; (6) logarithmic BH mass in $M_{\odot}$ units, from \citet{Vietri2018}.
\end{tablenotes}
\end{threeparttable}
\end{table*}

\begin{table*}
\centering
\begin{threeparttable}
\caption{Properties of the ASCA sample.} 
\label{tab:asca}
\begin{tabular}{llccc}
\hline
{Name} & $z$ & log$L_{X}$ & log$L_{BOL}$ & log$L_{O}$ \\
(1) & (2) & (3) & (4) & (5) \\

\hline
\hline
	&		\textbf{Type 2 sources} &	\\[2ex]

J235554+2836   &      0.729		&   44.87 &  46.36	&		-           \\
J130453+3548   &      0.316		&   43.91 &  44.79	&		-           \\
J090053+3856   &      0.229		&   43.94 &  44.54	&		-			\\
J234725+0053   &      0.213		&   43.93 &  45.19	&		-           \\
J111432+4055   &      0.153		&   43.77 &  44.25	&		-           \\
J150339+1016   &      0.095		&   43.26 &  43.80	&		-           \\
                                                                    
\hline                                                        
	& \textbf{Type 1 sources} &	\\[2ex]                             
	                                                                
J001913+1556    &     2.270 	& 45.71 &  46.76 &			46.54       \\
J015840+0347    &     0.658		& 44.75 &  46.35 &			45.36       \\
J125732+3543    &     0.524		& 44.56 &  45.69 &			45.21       \\
J002619+1050    &     0.474		& 44.24 &  45.50 &			45.02       \\
J023520-0347    &     0.376		& 44.23 &  45.65 &			44.89       \\
J151441+3650    &     0.371		& 44.83 &  45.96 &			45.30       \\
J144055+5204    &     0.320		& 44.19 &  45.43 &			44.89       \\
J000927-0438 	&     0.314		& 43.95 &  44.78 &			44.34	     \\
J121427+2936    &     0.309		& 44.25 &  45.03 &			44.90       \\
J122017+0641    &     0.287		& 44.33 &  45.27 &			44.79       \\
J142353+2247    &     0.282		& 44.25 &  45.12 &			44.44       \\
J172938+5230    &     0.278		& 44.13 &  45.41 &			44.98       \\
J103934+5330    &     0.229		& 43.62 &  44.61 &			44.12       \\
J233253+1513    &     0.215		& 44.13 &  44.98 &			44.54       \\
J121854+2957    &     0.178		& 43.64 &  45.16 &			43.63       \\
J170305+4526    &     0.171		& 43.68 &  44.97 &			44.50       \\
J140528+2224    &     0.156		& 43.44 &  44.52 &			44.11       \\
J121359+1404    &     0.154		& 43.26 &  44.36 &			43.92       \\
J170548+2412    &     0.114		& 43.26 &  44.60 &			43.90       \\
J134450+0005    &     0.087		& 42.95 &  44.25 &			43.91       \\
J121930+0643    &     0.081		& 43.00 &  44.52 &			44.14       \\
J144109+3520    &     0.077		& 42.64 &  43.95 &			43.66       \\
	
\hline

\end{tabular}

\begin{tablenotes}
\small
\item (1) Source name; (2) redshift; (3) intrinsic logarithmic hard X-ray luminosity (2-10 keV) in erg/s units from \citet{Liu2016} re-scaled for our adopted cosmology; (4) SED-fitting derived logarithmic bolometric luminosity in erg/s units; (5) B-band 4400 $\AA$ AGN logarithmic luminosity in erg/s units from the photometric data, provided only for type I sources, as the contamination from the galaxy is negligible.
\end{tablenotes}
\end{threeparttable}
\end{table*}

\begin{table*}
\centering
\begin{threeparttable}
\caption{Properties of the XXL sample.} 
\label{tab:xxl}
\begin{tabular}{llcccc}
\hline
{Name} & $z$ & log$L_{X}$ & log$L_{BOL}$ & log$L_{O}$ &  logM$_{BH}$\\
(1) & (2) & (3) & (4) & (5) & (6) \\

\hline
\hline

1237679323935212347		&		5.0111		&	45.24	&	47.10	&	46.38  &	9.42		\\
1237679253060517914     &		3.1977		&	44.86	&	46.71	&	45.82  &	8.60		\\
1237679253062091149     &		3.1737		&	45.12	&	46.54	&	45.75  &	8.24		\\
1237679322324271694     &		3.0009		&	44.84	&	46.58	&	45.51  &	8.19		\\
1237679323399389313     &		2.7648		&	45.57	&	46.70	&	46.08  &	9.41		\\
1237679439888187748     &		2.7638		&	44.82	&	46.24	&	-      &	8.10		\\
1237679324471623799     &		2.71  		&	45.16	&	46.94	&	46.27  &    9.11		\\
1237679253060780145     &		2.7074		&	44.81	&	46.64	&	45.97  &	9.43		\\
1237679323398013411     &		2.647 		&	45.02	&	46.44	&	45.95  &    8.76		\\
1237679254134259865     &		2.6351		&	44.93	&	47.07	&	46.44  &	9.81		\\
1237679323399061589     &		2.6016		&	44.97	&	47.06	&	46.40  &	9.70		\\
1237679322324992234     &		2.3271		&	45.02	&	46.31	&	45.59  &	8.76		\\
1237679322861601161     &		2.2212		&	44.76	&	46.07	&	45.23  &	7.95		\\
1237679322324336823     &		2.0902		&	44.70	&	46.50	&	45.94  &	8.77		\\
1237679321787662348     &		2.0375		&	45.09	&	47.08	&	-      &	9.93		\\
1237679254135505041     &		2.0092		&	45.14	&	46.41	&	45.85  &	9.09		\\
1237678887988167005     &		1.9572		&	44.58	&	46.27	&	45.36  &	7.91		\\
1237679254672113748     &		1.9153		&	44.74	&	46.25	&	45.64  &	8.85		\\
1237679323934621916     &		1.8757		&	44.96	&	46.89	&	46.01  &	9.24		\\
1237679322862256290     &		1.871 		&	45.29	&	47.22	&	46.67  &    9.91		\\
1237679253060714853     &		1.8508		&	44.37	&	46.17	&	45.38  &	8.36		\\
1237679322862321678     &		1.7039		&	44.97	&	46.97	&	46.41  &	9.59		\\
1237679255209115743     &		1.6757		&	44.90	&	46.53	&	45.92  &	8.95		\\
1237669770301735049     &		1.6423		&	44.56	&	46.28	&	45.65  &	8.39		\\
1237679253598044260     &		1.5449		&	44.58	&	46.59	&	45.90  &	9.34		\\
1237679253597782176     &		1.397 		&	44.59	&	46.73	&	46.04  &    9.33		\\
1237669770301669421     &		1.371 		&	44.81	&	46.62	&	46.01  &    9.39		\\
1237679255745986570     &		1.2249		&	44.53	&	46.69	&	46.02  &	9.19		\\
1237679323399127053     &		1.1815		&	44.25	&	46.68	&	46.12  &	9.27		\\
1237679253599092740     &		1.1644		&	44.65	&	46.63	&	45.84  &	8.87		\\
1237679322322632786     &		0.8909		&	44.29	&	46.09	&	45.53  &	8.30		\\
\hline
\end{tabular}
\begin{tablenotes}
\small
\item (1) Source name; (2) redshift; (3) intrinsic logarithmic hard X-ray luminosity (2-10 keV) in erg/s units; (4) SED-fitting derived logarithmic bolometric luminosity in erg/s units; (5) B-band 4400 \AA \, logarithmic luminosity in erg/s units from the photometric data when present; (6) logarithmic $M_{BH}$ in $M_\odot$ unit.
\end{tablenotes}
\end{threeparttable}
\end{table*}

\end{appendix}

\end{document}